\documentclass[doublespace]{revtex4-1}
\usepackage{amsmath, amsthm, amssymb, amsfonts, enumerate}
\usepackage{setspace}
\usepackage{natbib, bbm}
\usepackage{graphicx}
\usepackage{color}
\usepackage{longtable}
\usepackage{epstopdf}

\theoremstyle{plain}

\theoremstyle{definition}

\begin{document}

\title{ECOM: a fast and accurate solver for toroidal axisymmetric MHD equilibria}
\author{Jungpyo Lee}
\affiliation{Courant Institute of Mathematical Sciences, New York University, New York, NY 10012, USA}
\author{Antoine Cerfon}
\affiliation{Courant Institute of Mathematical Sciences, New York University, New York, NY 10012, USA}
\date{\today}

\begin{abstract}
We present ECOM (Equilibrium solver via COnformal Mapping), a fast and accurate fixed boundary solver for toroidally axisymmetric magnetohydrodynamic equilibria with or without a toroidal flow. ECOM combines conformal mapping and Fourier and integral equation methods on the unit disk to achieve exponential convergence for the poloidal flux function as well as its first and second partial derivatives. As a consequence of its high order accuracy, for dense grids and tokamak-like elongations ECOM computes key quantities such as the safety factor and the magnetic shear with higher accuracy than the finite element based code CHEASE [H. L\"utjens \textit{et al.}, Computer physics communications 97, 219 (1996)] at equal run time. ECOM has been developed to provide equilibrium quantities and details of the flux contour geometry as inputs to stability, wave propagation and transport codes.
\end{abstract}
\maketitle

\section{Introduction}\label{sec:introduction}

Numerically computed magnetohydrodynamic (MHD) equilibria are the starting point of a wide class of numerical solvers that are used to study MHD stability, transport, and heating and current drive in magnetic fusion devices \cite{sovinec,lapillonne,gorler,brambilla}. Static MHD equilibria of toroidally axisymmetric configurations are described by the Grad-Shafranov (G-S) equation \cite{grad1958hydromagnetic, shafranov1958}, a nonlinear, second-order elliptic partial differential equation. Stationary equilibria with purely toroidal flows are determined by solving a close variant of the G-S equation \cite{jardin2010computational}, the only difference being that for the latter the pressure term does not only depend on the poloidal flux function $\Psi$, but also on the radial variable $R$. Numerical codes to solve the G-S equation have been developed since the early days of the magnetic fusion program \cite{takeda1991computation, jardin2010computational,goedbloed2010advanced}. Nevertheless, the development of optimized G-S codes remains a topic of active research, for three main reasons. First, G-S solvers must be able to properly resolve complex two-dimensional geometries \cite{howell, li}, with boundaries that may have a corner, corresponding to a magnetic field X-point \cite{li}. Second, G-S solvers must be fast. This criterion is particularly relevant in the context of multiphysics integrated simulations \cite{candy,barnes,cary,chang,voitsekhovitch}. Several of these multiscale, multiphysics solvers already include, or will eventually include, in their iterative procedure a step in which the equilibrium configuration is self-consistently updated. A reasonable requirement is that the calculation of the updated equilibrium takes a negligible amount of time and computing resources as compared to the computationally intensive transport, MHD stability and plasma heating solvers. Third, G-S solvers must be accurate. The solution of the G-S equation is the poloidal flux $\Psi$, but the physical quantities of interest, such as the magnetic field, the safety factor, the magnetic shear, the magnetic curvature, and the current density are all functions of partial derivatives of $\Psi$. Since there always is some loss of accuracy when computing derivatives, a high level of accuracy for $\Psi$ is desired.

In this article, we present the new Grad-Shafranov code ECOM (Equilibrium solver via COnformal Mapping). ECOM is a fixed boundary, direct solver written in Fortran 77/90 that is based on three key elements: 1) the formulation of the G-S equation as a nonlinear Poisson problem; 2) a spectrally accurate numerical method to compute the conformal map from the smooth plasma cross section of interest to the unit disk; 3) a fast, high order Poisson solver on the unit disk \cite{pataki2013fast}. Its main novelty lies in the last two aspects discussed in the paragraph above, namely accuracy and speed. Regarding the first point, we demonstrate in this article that ECOM has better convergence properties than popular G-S solvers based on finite elements \cite{howell,lutjens1996chease,jardinfem}. In the finite element approach the numerical error of the solution decays as a power of grid size, i.e. $N^{-\alpha}$ where $\alpha$ is an integer and $N$ is the number of grid points in one direction. Often, for magnetic fusion applications $\alpha\leq4$ \cite{lutjens1996chease,goedbloed2010advanced}, although $\alpha\leq7$ was recently demonstrated \cite{howell}. In contrast, convergence in ECOM is exponential: the error decays as $\beta^{-N}$ for some real number $\beta>1$. Just as importantly, in ECOM the rate of convergence for the derivatives of $\Psi$ is the same as that of $\Psi$, whereas in the finite element approach the derivatives of $\Psi$ converge more slowly than $\Psi$ \cite{howell}. Remarkably, in ECOM numerical accuracy is not obtained at the expense of computational complexity and speed. For a given grid size, our solver is faster than finite element solvers and less demanding in terms of memory. We will show that a drawback of relying on conformal mapping is that ECOM often requires a denser grid than FEM based solvers to achieve a desired accuracy. Yet despite this, we find that for tokamak geometries and medium to high number of grid points, ECOM is more accurate than FEM based equilibrium codes at equal run time.

This article follows an earlier article \cite{pataki2013fast}, in which we gave a detailed description of our new numerical algorithm for solving the G-S equation. The focus here is different. One of the main motivations is to present extensions recently added to our G-S solver that make it a practical tool readily usable in fusion applications. The new capabilities of our equilibrium solver include the possibility of computing equilibria with arbitrary toroidal flow profiles, the possibility of specifying current and pressure profiles in various ways, as well as the evaluation of the key physical quantities that are required as inputs in stability, transport and heating codes. Since ECOM is a direct solver that calculates $\Psi$ on a prescribed grid for the poloidal cross section \cite{zakharov}, we put a particular emphasis on the accurate computation of the contours of constant flux, and of flux surface quantities such as the safety factor and the magnetic shear. A second motivation for this article is to perform detailed comparisons between the popular G-S code CHEASE \cite{lutjens1996chease} and ECOM, and to assess the merits of each solver.

The structure of the article is as follows. In Section \ref{sec:GS} we briefly review the numerical algorithm we use to solve the G-S equation \cite{pataki2013fast}. In Section \ref{sec:post}, we give a detailed presentation of the equilibrium quantities ECOM computes during the postprocessing phase, and of the numerical methods we implemented to calculate these quantitites with high accuracy.  In Section \ref{sec:Accuracy} we evaluate the speed, accuracy, and convergence properties of our solver, and compare them to those of CHEASE \cite{lutjens1996chease}. In Section \ref{sec:torflow} we explain how ECOM computes stationary equilibria with toroidal flows, and in Section \ref{sec:dis} we summarize our main findings, discuss the current limitations of ECOM and future plans. Appendix \ref{sec:Miller} presents our method to calculate the Miller parametrization \cite{miller} of a numerically computed flux contour, and Appendix \ref{sec:namelist} contains a table with all the important variables in ECOM, along with a short description for each of them.
 
\section{Numerical algorithm}\label{sec:GS}
In this section, we briefly review the numerical algorithm used in ECOM to solve the G-S equation. A more detailed presentation of each of the steps described below can be found in \cite{pataki2013fast}.

\subsection{The Grad-Shafranov equation as a nonlinear Poisson problem}
The Grad-Shafranov equation is given by
\begin{equation}
\Delta^{*}  \Psi\equiv R\frac{\partial}{\partial R}\left(\frac{1}{R}\frac{\partial\Psi}{\partial R}\right)  +\frac{\partial^2 \Psi}{\partial Z^2}=-\mu_{0} R^2\frac{d p(\Psi)}{d \Psi}- \frac{1}{2}\frac{d F^{2}(\Psi)}{d \Psi}\label{GS1}
\end{equation}
where $(R,\phi,Z)$ is the usual cylindrical coordinate system associated with the toroidal geometry, $2\pi\Psi$ is the poloidal magnetic flux, $\mu_{0}$ is the permeability of free space, $p(\Psi)$ is the plasma pressure, and $F(\Psi)=R B_\phi$, with $B_\phi$ the toroidal component of the magnetic field. Once the free functions $p(\Psi)$ and $F(\psi)$ are given and Eq. (\ref{GS1}) is solved with appropriate boundary conditions, the magnetic field $\mathbf{B}$ and the current density $\mathbf{J}$ can be computed according to the following formulae
\begin{displaymath}
\mathbf{B}=\frac{F(\Psi)}{R}\mathbf{e}_{\phi}+\frac{1}{R}\nabla\Psi\times\mathbf{e}_{\phi}\qquad\mathbf{J}=\frac{1}{\mu_{0}R}\frac{dF}{d\Psi}\nabla\Psi\times\mathbf{e}_{\phi}-\frac{1}{\mu_{0}R}\Delta^{*}\Psi\mathbf{e}_{\phi}
\end{displaymath}

Eq. (\ref{GS1}) is a second-order elliptic nonlinear partial differential equation for $\Psi$. ECOM solves the fixed boundary problem associated with this equation. Specifically, the boundary curve $\partial\Omega$ enclosing the plasma domain $\Omega$ of interest is an input to the solver, and ECOM solves Eq. (\ref{GS1}) with the Dirichlet data $\Psi=\Psi_{b}$ on $\partial\Omega$, where $\Psi_{b}$ is a constant. This formulation is particularly convenient for multiphysics theoretical studies of the influence of shaping on plasma performance \cite{marinoni,camenen,parra}. Two types of inputs can be used in ECOM to determine the geometry of $\partial\Omega$. One option is to give an exact representation of the plasma boundary, for example in the form of parametric equations \cite{miller}. When such a representation is not available, one can also give the coordinates $(R_{n},Z_{n})$ of discrete points on the boundary. At the moment, ECOM can only treat smooth plasma boundaries, and can therefore not compute equilibria whose plasma boundary has a separatrix.

The functional dependence on $\Psi$ of the pressure and toroidal magnetic field profiles is either prescribed or determined from transport equations. In both cases, it is an input to ECOM. In general, these profiles are such that Eq. (\ref{GS1}) is nonlinear, and for a wide class of profiles Eq. (\ref{GS1}) has to be solved as an eigenvalue problem \cite{takeda1991computation,pataki2013fast,goedbloed,lodestro}. This means that Eq. (\ref{GS1}) has to be solved by iterating on $\Psi$ \cite{trefethen}. In ECOM, this is done as follows. A normalized flux $\psi$ is defined by $\psi=(\Psi-\Psi_b)/(\Psi_0-\Psi_b)$, where $\Psi_0$ and $\Psi_B$ are the poloidal flux at the magnetic axis and the last closed flux surface respectively, so that $\psi=1$ at the magnetic axis and $\psi=0$ at the last closed flux surface. The pressure and toroidal magnetic field profiles are also normalized and expressed in terms of $\psi$ according to:
\begin{equation}\label{normal_prof}
\frac{dp(\Psi)}{d\Psi}=\frac{d\bar{p}(\psi)}{d\psi}\qquad\text{and}\qquad\frac{dF^{2}(\Psi)}{d\Psi}=\frac{d\bar{F}^{2}(\psi)}{d\psi}.
\end{equation}
Defining $\lambda=1/(\Psi_0-\Psi_b)$, Eq. (\ref{GS1}) then becomes
\begin{equation}
\Delta^{*}\psi = -\lambda\left(\mu_{0}R^{2}\frac{d\bar{p}}{d\psi}+\frac{1}{2}\frac{d\bar{F}^{2}}{d\psi}\right)\label{eigen}
\end{equation}
where $\lambda$ plays the role of an eigenvalue. In ECOM, there are several options to specify the profiles $d\bar{p}/d \psi$ and $d\bar{F}^{2}/d\psi$, with corresponding namelist parameter IPTYPE for the pressure and IFTYPE for the poloidal current. If IPTYPE=1 or  IFTYPE=1, the profiles are given by an explicit formula in terms of $\psi$. In ECOM, we often use $d\bar{p}/d \psi=p_{0\psi}(1-(1-\psi)^{p_{in}})^{p_{out}}$, as is also done in CHEASE \cite{lutjens1996chease}, where the constants $p_{0\psi}$, $p_{in}$, and ${p_{out}}$ are specified in the namelist. Likewise, we often use $d\bar{F}^{2}/d \psi=2F_{0\psi}(1-(1-\psi)^{F_{in}})^{F_{out}}$. Different expressions can be easily implemented, such as formulae describing a steep pressure pedestal \cite{pataki2013fast}. If IPTYPE=2 or IFTYPE=2, the profiles are given by a set of data points and the corresponding values of $\psi$ or of the minor radius. The value of the namelist variable IPTABLE determines whether the tabulated values of the profiles are in terms of $\psi$ or of the minor radius. If IPTABLE=0, the numerical tables of $d\bar{p}/d \psi$ and $d\bar{F}^{2}/d \psi$ are specified in terms of discrete values of $\psi$. If IPTABLE=1, $\bar{p}(\psi)$ and $\bar{F}(\psi)$ instead of their derivatives are specified by tables in terms of $\psi$. If IPTABLE=2, the numerical tables of $\bar{p}$ and $\bar{F}$ are specified in terms of the normalized minor radius $\rho$. In Section \ref{sec:post}, we explain in detail how ECOM accurately computes $d\bar{p}/d \psi$ and $d\bar{F}^{2}/d \psi$ starting from such tables. Furthermore, ECOM offers the possibility to choose between three different definitions for the minor radius $\rho$, corresponding to three different values of the namelist variable IRHO. If IRHO=0, the minor radius is defined by $\rho(\psi)=(R_o(\psi)-R_0)/(R_o(\psi=0)-R_0)$, where $R_o(\psi)$ is the outer location of the flux surface $\psi$ at $Z=0$ and $R_0$ is the location of the magnetic axis. If IRHO=1, $\rho$ is defined by $\rho(\psi)=({R_o(\psi)-R_i(\psi)})/({R_o(\psi=0)-R_i(\psi=0)})$ where $R_i(\psi)$ is the inner location of the flux surface $\psi$ at $Z=0$. If IRHO=2, $\rho$ is defined by $\rho=\sqrt{(\Psi-\Psi_0)/(\Psi_b-\Psi_0)}$.

As already mentioned, Eq. (\ref{eigen}) must be solved iteratively. G-S solvers usually iterate on the operator $\Delta^{*}$ \cite{howell,lutjens1996chease}. One of the key ideas in ECOM is to iterate on the Laplacian operator $\Delta$ instead. This can be done without any loss of generality, and does not incur any additional computational cost. The advantage is that at a given iteration, one now has to solve Poisson's equation, and one can therefore rely on the larger body of numerical methods developed for fast high order Poisson solvers in two dimensions. To solve for $\psi$ and the smallest eigenvalue $\lambda$ in Eq.(\ref{eigen}), ECOM uses a modified version of the inverse iteration method \cite{trefethen}. Specifically, if $\psi^{(i)}$ and $\lambda^{(i)}$ are known at the iteration step $i$, then $\psi^{(i+1)}$ and $\lambda^{(i+1)}$ are computed according to
\begin{align}
&\Delta\tilde{\psi}^{(i+1)}=\frac{1}{R}\frac{\partial \psi^{(i)}}{\partial R}-\lambda^{(i)}\left[\mu_{0}R^{2}\frac{d\bar{p}}{d\psi}(\psi^{(i)})+\frac{1}{2}\frac{dF^{2}}{d\psi}(\psi^{(i)})\right]\label{Poisson1}\\
&\psi^{(i+1)}=\frac{\tilde{\psi}^{(i+1)}}{||\tilde{\psi}^{(i+1)}||_{\infty}}\label{psiup}\\
&\lambda^{(i+1)}=\frac{\lambda_{i}}{||\tilde{\psi}^{(i+1)}||_{\infty}}.\label{lambdaup}
\end{align}
where $||\tilde{\psi}^{i+1}||_{\infty}$ is the extremum of $\tilde{\psi}^{i+1}$ in the domain $\Omega$. In ECOM, the iterative process terminates when $||\psi^{(i+1)}-\psi^{(i)}||_{\infty}<\delta$ for some predetermined small $\delta$. In the numerical results presented in this article, we typically had $\delta=10^{-14}$. In order to keep consistency across different grid sizes, $||\tilde{\psi}^{i+1}||_{\infty}$ is not only calculated for all values of $\tilde{\psi}^{i+1}$ on the mesh, but instead over the entire domain $\Omega$. In ECOM, this is done by finding the location where $\nabla\psi=0$ with the Newton-Raphson method. It requires knowledge of the Hessian matrix, whose values away from grid points are evaluated by interpolation, based on the high order Fourier and Chebyshev representations ECOM uses for the Poisson solver on the unit disk. The extremum of $\psi$ on the grid is used to provide a very good initial guess, so that in practice very few Newton steps are subsequently required to find $||\tilde{\psi}^{(i+1)}||_{\infty}$ in $\Omega$.

It is known empirically that iterative schemes such as the one above converge faster when the right-hand side of the partial differential equation is slowly varying. For better convergence we thus scale the unknown function $\psi$ as $\psi=u\sqrt{R}$. Replacing $\psi$ with $u$, Eq.(\ref{Poisson1}) becomes
\begin{equation}
\Delta\tilde{u}^{(i+1)}=\frac{3}{4}\frac{u^{(i)}}{R^{2}}-\frac{\lambda^{(i)}}{\sqrt{R}}\left[\mu_{0}R^{2}\frac{d\bar{p}}{d(\sqrt{R}u)}(\sqrt{R}u^{(i)})+\frac{1}{2}\frac{dF^{2}}{d(\sqrt{R}u)}(\sqrt{R}u^{(i)})\right]\label{Poisson2}
\end{equation}

Unlike Eq. (\ref{Poisson1}), the right-hand side of Eq. (\ref{Poisson2}) does not have derivatives of $\psi$, and it is therefore smoother. Note that while one solves for $u$ in the Poisson step, the normalization steps (\ref{psiup}) and (\ref{lambdaup}) are still computed in terms of $\psi$.

Thus far, we have explained how ECOM treats the Grad-Shafranov equation as a nonlinear Poisson problem. We now briefly describe how ECOM computes the solution of the Poisson equation (\ref{Poisson2}) at each iteration, i.e. with fixed right-hand side, on a domain $\Omega$ of fusion interest and with the Dirichlet boundary condition $\tilde{u}=0$ on $\partial\Omega$. The Poisson solver in ECOM is based on two elements: 1) a spectrally accurate numerical method to compute the conformal map from the plasma domain $\Omega$ to the unit disk; 2) a fast, high order Poisson solver on the unit disk. 

\subsection{Conformal maping from the plasma domain to the unit disk}
Conformal mapping is an effective method for solving Poisson's equation because a conformal map transforms a Laplacian operator into another Laplacian operator, with a scale factor \cite{goedbloed2010advanced}. Consider the generic Poisson equation
\begin{equation}\label{Poisson_generic}
\begin{cases}
\Delta u(R,Z) = f(R,Z)\qquad\mbox{in}\;\Omega\\
u(R,Z)=0\qquad\mbox{on}\;\partial\Omega
\end{cases}
\end{equation}
the conformal map $W:z=R+iZ\mapsto w=\alpha+i\beta$ from $\Omega$ to the unit disc $D_{1}$ and its inverse map $B:w=\alpha+i\beta\in D_{1}\mapsto z=R+iZ\in\Omega$. Solving Eq. (\ref{Poisson_generic}) is equivalent to solving the following Poisson problem in $D_{1}$:
\begin{equation}\label{Poisson_mapped}
\begin{cases}
\Delta v(\alpha,\beta) = f(R(\alpha,\beta),Z(\alpha,\beta))\left|\frac{dB}{dw}\right|^{2}\qquad\mbox{in}\;D_{1}\\
v(\alpha,\beta)=0\qquad\mbox{on}\;\partial D_{1}
\end{cases}
\end{equation}
where $u(R,Z)=v(\alpha(R,Z),\beta(R,Z))$, and the functions $R(\alpha,\beta)$ and $Z(\alpha,\beta)$ should be seen as the real and imaginary parts of the inverse map $B$. Clearly, solving Eq. (\ref{Poisson_mapped}) is easier than solving Eq. (\ref{Poisson_generic}), provided one has a way to calculate the inverse map $B$ at both boundary and interior points of $D_{1}$. In ECOM, this is done as follows. ECOM first computes the forward map $W$ for points on $\partial\Omega$ that are equispaced in arc length through the Kerzman-Stein integral equation based on the Szeg\"o kernel \cite{pataki2013fast,kerzman_stein,kerzman_trummer}. Using oversampling and interpolation, ECOM then uses the boundary values of the forward map to calculate the inverse map $R(\alpha,\beta)$ and $Z(\alpha,\beta)$ for points on the boundary of $D_{1}$ that are equispaced in the polar angle $\vartheta$. Finally, ECOM computes $B$ for points in the interior of $D_{1}$ using the Cauchy integral formula and the Fast Fourier Transform \cite{pataki2013fast}.

ECOM relies on a somewhat naive implementation of the Kerzmann-Stein integral equation for the computation of $W$ on $\partial\Omega$, that requires $O(n_{1}^{3})$ work, where $n_{1}$ is the number of discretization points on $\partial\Omega$. There exist methods resulting in an asymptotic $O(n_{1})$ run time \cite{andras_thesis}, but they are not currently implemented in ECOM because the computation only needs to be done once, and because other steps in ECOM are more expensive. The computation of the inverse map at interior points is based on the Cauchy integral formula and the Fast Fourier Transform, and results in a run time complexity of $O(n_{r}n_{\vartheta}\mbox{log}n_{\vartheta})$, where $n_{r}$ is the number of radial grid points and $n_{\vartheta}$ the number of angular grid points for the mesh that ECOM uses to solve Poisson's equation on $D_{1}$ 

The boundary $\partial\Omega$ can be defined in several ways, that are specified by the namelist variable IBTYPE in ECOM.  If IBTYPE=0, the plasma boundary corresponds to the contour $\Psi=0$ of a Solov'ev equilibrium we discuss in Section \ref{sec:Solov'ev}, parametrized by Eqs. (\ref{solob1})-(\ref{solob2}). If IBTYPE=1, the boundary is specified by the Miller parametrization \cite{miller} given by Eqs. (\ref{millerb1})-(\ref{millerb2}) of Section \ref{sec:exnl}, and the elongation $\kappa$ and triangularity $\delta_{m}$ must then be specified. If IBTYPE=2, the boundary is specified by a set of discrete points $(R,Z)$. This allows the computation of equilibria specified by experimental data and is also the method of choice to compute up-down asymmetric equilibria. The conformal mapping routine requires that the points on $\partial\Omega$ be equispaced in arc length. When IBTYPE=0 or IBTYPE=1, these points are easily calculated from the parametric equations for the boundary. When IBTYPE=2, ECOM uses Lagrange interpolation to compute these points.   
 
In principle, the conformal map has to be computed only once, at the beginning of the iterative procedure corresponding to Eqs. (\ref{psiup})-(\ref{Poisson2}). However, as one might physically expect, computing flux surface quantities is much more convenient if the point that is mapped to the center of $D_{1}$ coincides with the magnetic axis. To facilitate the calculation of these quantities, which takes place after the G-S equation is solved, ECOM recomputes the conformal map several times within the Poisson iterations to adjust the center of $D_{1}$ to the magnetic axis. We have empirically observed that the conformal map only needs to be recomputed a few times.

The Riemann mapping theorem guarantees the existence and uniqueness of an analytic map between any simply connected plasma cross section $\Omega$ and the unit disk. However, this does not mean that conformal mapping is a practical numerical method for any arbitrary plasma shape. The issue is that points on the boundary of $D_{1}$ that are equispaced in the angle $\vartheta$ are not necessarily mapped, under the inverse map, to points that resolve the boundary of $\Omega$ in the desired fashion. It is well known, for example, that if $\Omega$ is an elongated ellipse, uniformly spaced points in $\vartheta$ on the boundary of $D_{1}$ correspond to a distribution of points on the boundary of $\Omega$ which is sparse on the curved parts and crowded on the flat parts of the ellipse \cite{goedbloed2010advanced, pataki2013fast}. In the remainder of this article, we will call this phenomenon the ``crowding effect". It has two direct implications for ECOM. First, ECOM can only treat in a robust manner domains that have a smooth boundary, and can therefore not be used to compute equilibria with a magnetic X-point on the plasma boundary. Second, ECOM is particularly efficient for plasma shapes that are not too elongated, as is the case for conventional tokamaks. As elongation is increased, the high order convergence properties are maintained for $\Psi$ and its derivatives, but a higher number of grid points is required to reach a certain level of accuracy \cite{pataki2013fast}. We will go back to this point in Section \ref{sec:dis}.

\subsection{Fast, high order Poisson solver on the unit disk} \label{sec:poisson}
We finish this section by describing how ECOM solves Poisson's equation on the unit disk,
\begin{equation}\label{Poisson_disk}
\begin{cases}
\Delta v = g\qquad\mbox{in}\;D_{1}\\
v(\alpha,\beta)=0\qquad\mbox{on}\;\partial D_{1}
\end{cases}
\end{equation}
as required in Eq. (\ref{Poisson_mapped}). The solver uses separation of variables in the usual polar coordinates $(r,\vartheta)$ and expands $v$ and $g$ as Fourier series
\begin{displaymath}
v(r,\vartheta)=\sum_{n=-\infty}^{\infty}\hat{v}_{n}(r)e^{in\vartheta}\qquad g(r,\vartheta)=\sum_{n=-\infty}^{\infty}\hat{g}_{n}(r)e^{in\vartheta}
\end{displaymath}
Substituting these expressions into Poisson's equation, we get the following ordinary differential equation for each $n$:
\begin{equation}\label{ode}
\begin{cases}
\hat{v}_{0}^{''}(r)+\frac{1}{r}\hat{v}_{0}^{'}(r)=\hat{g}_{0}(r)\qquad \hat{v}_{0}^{'}(0)=0\qquad \hat{v}_{0}(1)=0\\
\hat{v}_{n}^{''}(r)+\frac{1}{r}\hat{v}_{n}^{'}(r)-\frac{n^{2}}{r^{2}}\hat{v}_{n}(r)=\hat{g}_{n}(r)\qquad \hat{v}_{n}(0)=0\qquad \hat{v}_{n}(1)=0\qquad n\neq0
\end{cases}
\end{equation}
where the boundary condition at $r=0$ is obtained by requiring the regularity of the solution at this point. For each $n$, a solution of Eq. (\ref{ode}) that does not satisfy the boundary condition at $r=1$ can be written in terms of convolutions with the Green's function associated with Eq. (\ref{ode}) that has the proper behavior at $r=0$ and $r\rightarrow\infty$:
\begin{equation}\label{particular}
\begin{cases}
\hat{v}_{0}^{P}(r)=\mbox{log}r\int_{0}^{r}s\hat{g}_{0}sds+\int_{r}^{1}s\mbox{log}s\hat{g}_{0}(s)ds\\
\hat{v}_{n}^{P}(r)=-\frac{1}{2|n|}\left(r^{-|n|}\int_{0}^{r}s^{|n|+1}\hat{g}_{n}(s)ds+r^{|n|}\int_{r}^{1}s^{-|n|+1}\hat{g}_{n}(s)ds\right)\qquad n\neq0
\end{cases}
\end{equation}
For each $n$, the general solution to the homogeneous equation
\begin{equation}\label{homoeq}
\hat{v}_{n}^{''}(r)+\frac{1}{r}\hat{v}_{n}^{'}(r)-\frac{n^{2}}{r^{2}}\hat{v}_{n}(r)=0
\end{equation}
satisfying the regularity condition at $r=0$ can also be written explicitly:
\begin{equation}\label{homosol}
\hat{v}^{H}_{n}(r)=c_{n}r^{|n|}
\end{equation}
where $c_{n}$ is a constant to be determined from the boundary condition at $r=1$. Setting $c_{n}=-\hat{v}^{P}_{n}(1)$, we can then write the solution of Eq. (\ref{ode}) satisfying the proper boundary conditions as follows:
\begin{equation}\label{radial_sol}
\hat{v}_{n}(r)=\hat{v}^{P}_{n}(r)-\hat{v}^{P}_{n}(1)r^{|n|}
\end{equation}

One of the major advantages of the Green's function formulation and the formula (\ref{radial_sol}) used in ECOM is that partial derivatives of $v$ can be calculated explicitly from the formulae in Eqs. (\ref{particular}) and (\ref{homosol}) \cite{pataki2013fast}. Numerical differentiation is never required, which is one of the main reasons why our numerical method leads to partial derivatives of $\psi$ that have the same order of convergence as $\psi$.

In ECOM, the computation and sum of the Fourier series are done with the Fast Fourier Transform. The angular grid on the unit disk is uniformly spaced in the polar angle $\vartheta$, to guarantee the spectral accuracy of the representation for smooth data. The number of grid points in the $\vartheta$ direction is $n_{\vartheta}$. In the radial direction, ECOM uses a piecewise Chebyshev grid. Specifically, the interval $[0,1]$ is divided into $n_{L}$ subintervals, and on each of these subintervals a Chebyshev grid of order $n_{ch}$ is constructed. The number of grid points in the radial direction is then $n_{r}=n_{ch}n_{L}$. $n_{ch}=16$ is the default setting in ECOM. The convolutions with the Green's function in Eq. (\ref{particular}) are computed with a 16th order Gaussian quadrature rule. Reference \cite{pataki2013fast} describes ways to avoid the computational issues associated with the rapid growth and decay of the monomials $s^{|n|}$ and $s^{-|n|}$ for large $n$, as well as the recursive algorithm used to compute these integrals in $O(n_{r})$ work. The run time complexity of the Poisson solver on the disk is $O(n_{r}n_{\vartheta}\mbox{log}n_{\vartheta})$: ECOM computes $O(n_{r})$ FFTs of size $n_{\vartheta}$ at a cost of $O(n_{r}n_{\vartheta}\mbox{log}n_{\vartheta})$ and solves $n_{\vartheta}$ radial ordinary differential equations at a cost of $O(n_{r}n_{\vartheta})$.

\begin{figure*}
\includegraphics[scale=0.7]{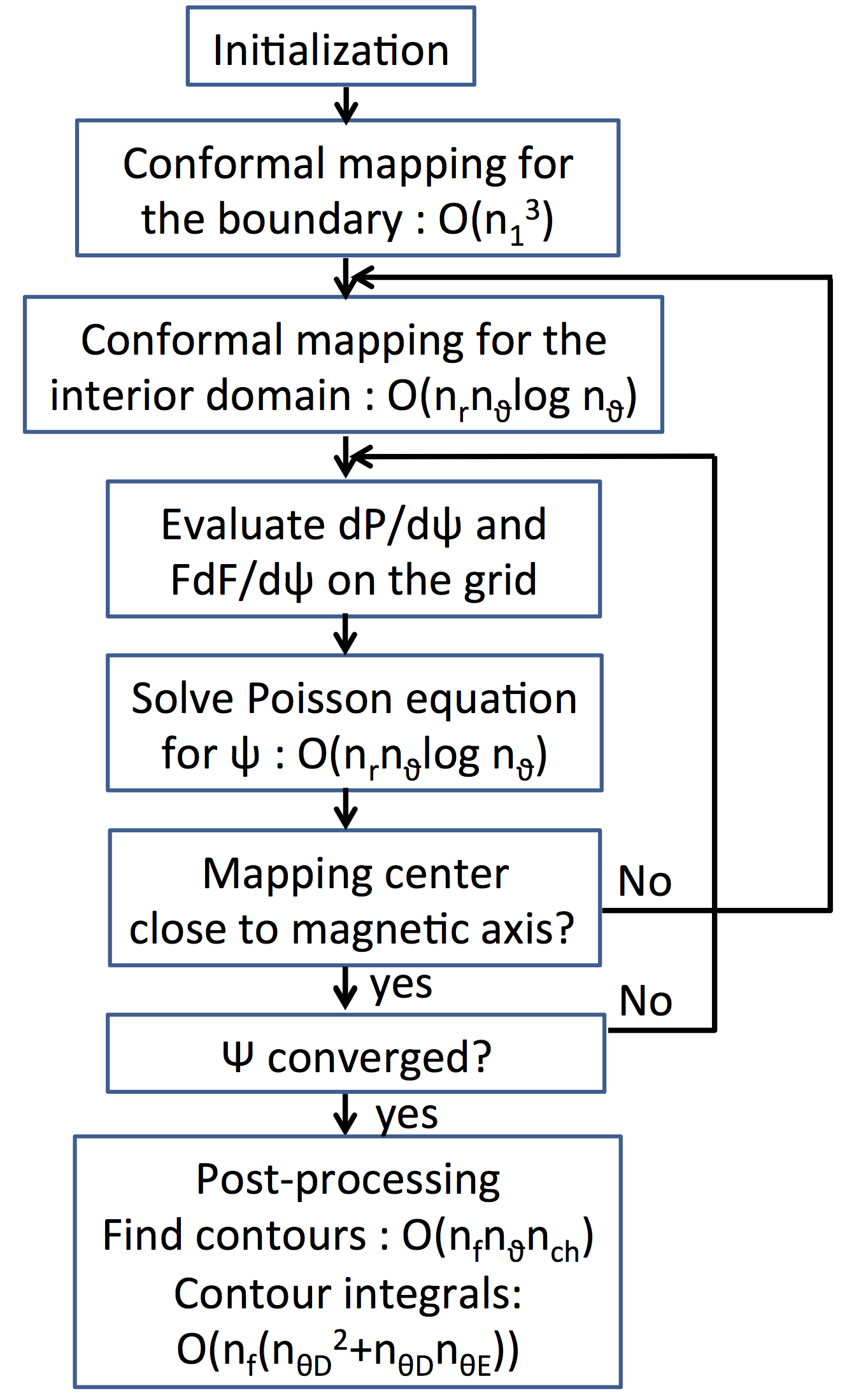}
\caption{Flow chart of ECOM code. The orders in the parenthesis indicate the run time complexity of the algorithm used for a given step}\label{ECOMdiagram}
\end{figure*}

The flow diagram in Figure \ref{ECOMdiagram} presents a condensed view of the iterative scheme used in ECOM, as a summary of Section \ref{sec:GS}. The initialization step corresponds to the specification of the grid resolution and of the values of the namelist variables presented in this article. The parameters $n_{f}$, $n_{\theta D}$ and $n_{\theta E}$ in Figure \ref{ECOMdiagram} refer to discretizations used during postprocessing, after the G-S equation is solved, and are defined in Section \ref{sec:post}. 

\section{Postprocessing}\label{sec:post}
The main purpose of postprocessing is to compute equilibrium quantities that play a key role in heating and current drive, stability and transport calculations, and to scale the normalized solution $\psi$ to the physical flux $\Psi$. In addition, ECOM can compute the parallel current density using reduced models for the Ohmic current and the bootstrap current, check the Mercier criterion for each flux surface \cite{Mercier} and the global Troyon limit \cite{Troyon}, and calculate the Miller parametrization that best fits a flux surface chosen by the user.

\subsection{Flux functions}\label{sec:post1}
The evaluation of flux quantities requires integration along contours of constant poloidal magnetic flux $\psi$. A flux coordinate system $(\psi,\theta,\phi)$ is more convenient for such integrals than the $(R, \phi, Z)$ coordinate system used by ECOM to solve the G-S equation. For postprocessing, ECOM thus constructs the following flux coordinate system. $\phi$ is chosen to be the usual toroidal angle $\phi$, with $\nabla \phi =\mathbf{e}_{\phi}/R$, while the poloidal angle $\theta$ is defined by $\theta=\mbox{atan2}(Z-Z_0,R-R_0)$ if $\mbox{atan2}(Z-Z_0,R-R_0)\geq0$, $\theta=\mbox{atan2}(Z-Z_0,R-R_0)+\pi$ if $\mbox{atan2}(Z-Z_0,R-R_0)<0$, where $\mbox{atan2}$ is the four-quadrant inverse tangent and $(R_0, Z_0)$ is the position of the magnetic axis. In terms of the $(\psi,\theta,\phi)$ coordinates, the flux surface average of a function $X$ \cite{jardin2010computational}, written $\langle X\rangle$, is defined by
\begin{equation}
\langle X \rangle=\frac{\int_0^{2\pi} d\theta JX}{\int_0^{2\pi}  d\theta J}
\end{equation}
where $J=(\nabla\psi\times\nabla\theta\cdot\nabla\phi)^{-1}$ is the Jacobian of the transformation between Cartesian coordinates and the $(\psi,\theta,\phi)$ coordinate system:
\begin{eqnarray}
\frac{1}{J}=\frac{1}{R}\left(\frac{\partial \psi}{\partial Z}\frac{\partial \theta}{\partial R}-\frac{\partial \psi}{\partial R}\frac{\partial \theta}{\partial Z}\right)
=-\frac{1}{R[(Z-Z_0)^2+(R-R_0)^2]}\left[(R-R_0)\frac{\partial \psi}{\partial R}+(Z-Z_0)\frac{\partial \psi}{\partial Z}\right]
\end{eqnarray}
As will soon be apparent, several physical quantities are best expressed in terms of the three functions $I_{A}(\psi)$, $I_{B}(\psi)$, and $I_{C}(\psi)$ defined by
\begin{equation}
 I_A(\psi)=\int_0^{2\pi} d\theta \frac{J |\nabla \psi|^2}{R^2},\qquad
  I_B(\psi)=\int_0^{2\pi} d\theta J ,\qquad 
   I_C(\psi)=\int_0^{2\pi} d\theta \frac{J}{R^2}.
 \label{int123}
\end{equation}
Rewriting Eq. (\ref{eigen}) using flux coordinates, a simple relationship can be obtained that links $I_{A}$, $I_{B}$, and $I_{C}$. Indeed, Eq. (\ref{eigen}) takes the following form in flux coordinates \cite{jardin2010computational}:
\begin{equation}\label{eq:GS_flux}
\frac{R^{2}}{J}\frac{\partial}{\partial\psi}\left(\frac{J}{R^{2}}|\nabla\psi|^{2}\right)+\frac{\partial}{\partial\theta}\left(\frac{J}{R^{2}}\nabla\theta\cdot\nabla\psi\right)=-\lambda\left(\mu_{0}R^{2}\frac{d\bar{p}}{d\psi}+\frac{1}{2}\frac{d\bar{F}^{2}}{d\psi}\right)
\end{equation}
Taking the flux surface average of $1/R^2 \times$ Eq. (\ref{eq:GS_flux}), we obtain the desired relationship between $I_{A}$, $I_{B}$, and $I_{C}$:
\begin{equation}
\frac{d I_A (\psi)}{d \psi}=-\lambda\left(\mu_{0} \frac{d \bar{p}}{d \psi}I_B+\frac{1}{2}\frac{d {\bar {F}^2}}{d \psi} I_C\right)\label{GSrsec3}.
\end{equation}
The plasma volume inside the flux surface $\psi$ is
\begin{eqnarray}
V(\psi)=2\pi\int_{1}^{\psi}d\tilde{\psi}\int_0^{2\pi} d\theta J= 2\pi\int_{1}^{\psi} d\tilde{\psi} I_B (\tilde{\psi}),\label{vol}
\end{eqnarray}
and the total plasma volume is $V_0=V(\psi=0)$. The toroidal current within the flux surface $\psi$ is
\begin{eqnarray}
I_\phi(\psi)=\int_{1}^{\psi}d\tilde{\psi}\int_0^{2\pi} d\theta j_\phi  \frac{J}{R} = -\int_{\psi}^{1}d\tilde{\psi}\left(\frac{ I_c}{2\mu_{0}}\frac{d {\bar {F}^2}}{d \tilde{\psi}}+ I_B\frac{d \bar{p}}{d \tilde{\psi}}\right)=\frac{1}{\lambda\mu_{0}}\int_{\psi}^{1}d\tilde{\psi}\frac{d I_A}{d \tilde{\psi}}=\frac{I_A(\psi)}{\lambda\mu_{0}},\label{torcur1}
\end{eqnarray}
where we have used Eq. (\ref{GSrsec3}) and $I_A(\psi=1)=0$. The total toroidal current in $\Omega$ is $I_{p}=I_\phi(\psi=0)$. The volume averaged pressure $\langle p \rangle_V$ is given by
\begin{equation}
\langle p \rangle_V=2\pi\frac{\int_{1}^{0}d\psi I_B(\psi)p(\Psi)}{V_{0}}=\frac{1}{\lambda V_0}\int_{0}^{1}d\psi V(\psi)\frac{d\bar{p}}{d\psi}\label{pres},
\end{equation}
where we have used integration by part, combined with $dV/d\psi=2\pi I_{B}$, $V(\psi=1)=0$, and $p(\psi=0)=0$. The total beta is 
\begin{equation}
\beta=\frac{2\mu_{0}\langle p \rangle_V}{ B_0^2},
\label{beta}
\end{equation}
where $B_0=F(\Psi=\Psi_B)/R_{mid}$ is the vacuum field at the point $(R_{mid},0)$, with $R_{mid}=(R_{i}(0)+R_{o}(0))/2$. Our definitions of the poloidal beta and of the internal inductance are the same as Jardin's \cite{jardin2010computational}:
\begin{align}
\beta_P&=\frac{4V_{0}\langle p \rangle_V}{\mu_{0}R_{0}I_{p}^{2}}
\label{betap}\\
l_i(\psi)&=\frac{4\pi}{\mu_{0}^{2}I_{p}^2R_0}\int_{1}^{\psi}d\psi \int_0^{2\pi} d\theta J\frac{|\nabla \Psi|^2}{R^2}=4\pi\frac{ \int_{1}^{\psi}d\psi I_A(\psi)}{I_A^2(\psi=0)R_0}.\label{intind}
\end{align}
The total poloidal magnetic field energy is 
\begin{eqnarray}
W_p= \pi\int_{1}^{0}d\psi\int_0^{2\pi} d\theta J\frac{|\nabla \Psi|^2}{R^2}=\frac{\pi}{\lambda^2}{\int_{1}^{0}d\psi I_A(\psi)}.\label{polE}
\end{eqnarray}
Finally, for each flux surface, the safety factor is defined as  
\begin{eqnarray}
q(\psi)=\lambda\frac{F[\Psi(\psi)]}{2\pi}\int_0^{2\pi} d\theta \frac{J}{R^2}=\lambda\frac{F[\Psi(\psi)]}{2\pi} I_C\label{q1}.
\end{eqnarray}
The normalized radius can be used to find the differential volume $dV(\psi)/d\rho$ and the differential flux $d\psi/d\rho$, which are often used in transport or MHD analysis. Also, the magnetic shear can be defined in terms of $\rho$ by
\begin{eqnarray}
\hat{s}(\psi)&=&\frac{\rho}{q(\psi)}\frac{dq(\psi)}{d\rho}. \label{shat}
\end{eqnarray}

\subsection{Numerical method for contour integrals}\label{sec:post2}
To evaluate the integrals in Eq. (\ref{int123}) numerically, one needs to find the location of the desired flux contours, which in general do not coincide with the $(R,Z)$ grid of the Poisson solver, and then integrate the integrands along these contours. In ECOM, this is done with the following three steps: (i) ECOM first determines the radial location of $n_{f}$ flux surfaces for each angle $\vartheta$ in the unit disk $D_1$; (ii) ECOM then evaluates the integrands at the corresponding $(R,Z)$ points in $\Omega$, and interpolates the integrands to a grid that is equispaced in the angle $\theta$; (iii) ECOM finally computes the integrals along contours of constant $\psi$ in the domain $\Omega$ using the trapezoidal rule. Since the numerical methods that are used for interpolation and integration in steps (i)-(iii) are spectrally accurate, ECOM computes the location of the flux contours and the integrals in Eq. (\ref{int123}) without significant loss of accuracy, as we will demonstrate in Section \ref{sec:Accuracy}. We now describe steps (i), (ii), and (iii) in more detail.

For a description of step (i), we put ourselves in the situation in which the G-S solver has computed the values $\psi(r_{i},\vartheta_{j})$ of the flux $\psi$ on the grid of the unit disk $D_{1}$, and we imagine that we want to determine the location of the contour $\psi=\psi_{s}$ on $D_{1}$. ECOM does this as follows. For each angle $\vartheta_{j}$, ECOM first finds the Chebyshev subinterval of the radial grid for which $\psi(r_{t},\varphi_{j})>\psi_{s}>\psi(r_{t+ch-1},\varphi_{j})$, where $t$ is the index of the first Chebyshev point in that radial subinterval. Once the subinterval is found, ECOM constructs a local continuous approximation $\psi_{c}(r,\vartheta_{j})$ of $\psi$ in the radial direction from the known values $\psi(r_i,\vartheta_j)$ on the subinterval of interest and the Chebyshev grid for that subinterval. Specifically, $\psi_{c}$ is written as the following sum: 
\begin{equation}\label{eq:Cheby_rep}
\psi_c(r,\vartheta_j)=\sum_{k=0}^{n_{ch}-1}a_kT_k(r)
\end{equation}
where the functions $T_{k}$ are the Chebyshev polymials associated with the Chebyshev grid of the subinterval, and where the coefficients $a_{k}$ are given by the expression
\begin{equation}
 a_k=\frac{2-\delta_{0k}}{n_{ch}}\sum_{p=0}^{n_{ch}-1} T_p(r_{t+p}) \psi(r_{t+p},\vartheta_j),
\end{equation}
where $\delta_{ij}$ is the kronecker delta.
ECOM then uses the expansion in Eq. (\ref{eq:Cheby_rep}) to find the radial position satisfying $|\psi_c(r)-\psi_s | < \delta$ with a Newton-Raphson iterative method:
 \begin{eqnarray}
 r^{q+1}=r^{q}-\frac{\psi_c(r^{q})-\psi_s}{\partial\psi_c/\partial r |_{r=r^{q}}} \;\;\;\;\;\;q=1,2,...,m.
 \end{eqnarray}
This root finding process usually converges in a few iterations, typically ${m}\leq 5$, and the total cost to find the location of $n_f$ contours at $n_{\vartheta}$ angles is $O(n_f n_\vartheta n_{ch})$. Finally, after the radial position of a given contour is found, ECOM uses $(\partial \psi/\partial r)(r_i, \vartheta_j)$ and $(\partial \psi/\partial \vartheta)(r_i, \vartheta_j)$ for $i=t, t+1, ..., t+n_{ch}-1$ and Chebyshev representations analogous to Eq. (\ref{eq:Cheby_rep}) to accurately evaluate $(\partial \psi/\partial r)(r, \vartheta_j)$ and $(\partial \psi/\partial \vartheta)(r, \vartheta_j)$ at the location of the flux contours. At the end of step (i) the radial position of the specified flux contours are known for each angle $\vartheta_{j}$ in $D_{1}$, and so are the values of the integrands in Eq. (\ref{int123}) at these points. Through the backward map, all these quantities are also known in $\Omega$.
 
In Eq. (\ref{int123}), one integrates quantities that are $2\pi$-periodic in $\theta$ over the period $[0,2\pi]$. Numerically, this can be done very accurately with a trapezoidal-rule quadrature, provided that the integrands are known on an equispaced $\theta$ grid. Since the equispaced $\vartheta$ grid in $D_{1}$ is not mapped to an equispaced $\theta$ grid, the goal of step (ii) is to interpolate the quantities computed in step (i) on a grid in $\Omega$ that is equispaced in $\theta$. Because of the crowding effect that is inherent to the conformal mapping technique, the angular grid resulting from the inverse map underresolves certain regions of $\Omega$. For accurate interpolation, it is thus desirable to first oversample the integrands to be interpolated. The oversampling is done by refining the equispaced $\vartheta$ grid of $D_{1}$ using the FFT, assuming band-limited integrands. As a result of this, the integrands are known at $n_\vartheta k_{samp}$ angular grid points, where $k_{samp}$ is the oversampling factor. In principle, these values could then be mapped back to $\Omega$, and interpolated. In practice, however, the backward mapping of the $n_\vartheta k_{samp}$ grid points for $n_{f}$ contours from $D_1$ to $\Omega$ requires $O(n_f  (n_\vartheta k_{samp})^2)$ operations, and results in significant computational time when $k_{samp}$ is as large as desired for accurate interpolation. Note that the run-time complexity is not of the form $n_f  n_\vartheta k_{samp}\mbox{log}(n_\vartheta k_{samp})$ as it was for the backward mapping of the $(r,\vartheta)$ grid points in $D_{1}$ because the calculation of the backward map for the flux contours cannot be trivially accelerated by the FFT. The reason for this is that the radial location of the flux contours varies as a function of $\vartheta$. In order to reduce the computational cost, we only compute the backward map for $n_{\theta D}$ points among the $n_\vartheta k_{samp}$ oversampled points, with $n_{\theta D}\ll n_\vartheta k_{samp}$, and chosen such that their mapped positions are the closest to the target grid of $n_{\theta E}$ points equispaced in $\theta$. These $n_{\theta D}$ points are found by computing the backward map of a small number of points lying on the contour of interest in $D_{1}$ with increasing angle $\vartheta$ across the interval $[0,2\pi]$, and using $k_{Lag}$-th order Lagrange interpolation to construct an approximation of the function $\vartheta(\theta)$ on that contour. We usually take $n_{\theta D}=n_{\theta E}$ and $k_{Lag}=8\ll k_{samp}$. 

Once the $n_{\theta D}$ non-equispaced points are found, we use trigonometric interpolation for a periodic function \cite{Berrut_barycentriclagrange} to interpolate the integrands at the $n_{\theta D}$ points to the $n_{\theta E}$ equispaced points. This requires $O(n_f  n_{\theta D}^2 )$ work to find the barycentric factors, and $O(n_f  n_{\theta D}n_{\theta E} )$ work to interpolate at $n_{\theta E}$ points. 
 
Once step (ii) is completed, step (iii) is straightforward. The contour integrals in Eq. (\ref{int123}) are computed from the integrands on the equispaced $\theta$ grid using the trapezoidal rule. Since the integrands are smooth and periodic in $\theta$, and since the $\theta$ grid is uniform, the trapezoidal rule is spectrally accurate \cite{trefethen2014exponentially}. The required work for the trapezoidal-rule quadrature is very small, $O(n_f  n_{\theta E})$.

\subsection{Scaling the equilibrium}\label{sec:post3}
A single solution of the normalized form of the G-S equation as given in Eq. (\ref{eigen}) can describe an infinite sequence of axisymmetric equilibria that have a different total toroidal current $I_{p}$ and a different safety factor $q_{0}$ at the magnetic axis. To understand these degrees of freedom, consider that ECOM has just computed the eigenvector-eigenvalue solution $(\psi,\lambda)$ of Eq. (\ref{eigen}). The normalized total toroidal current $I_{p}^{N}$ can be calculated from this solution according to Eq. (\ref{torcur1}). All is then needed to obtain an equilibrium with the desired total toroidal current $I_{p}^{D}$ is the simple rescaling $\lambda\rightarrow (I_{p}^{N}/I_{p}^{D})\lambda$, which is equivalent to rescaling $\Psi$. From the definitions in Eq. (\ref{normal_prof}), it is clear that $d\bar{p}/d\psi$ and $d\bar{F}^{2}/d\psi$ are also scaled by $\lambda$, and must be rescaled as well: $d\bar{p}/d \psi\rightarrow(I_{p}^{D}/I_{p}^{N})d\bar{p}/d\psi$ and $d\bar{F}^2/d\psi\rightarrow(I_{p}^{D}/I_{p}^{N})d\bar{F}^2/d\psi$. Once $\lambda$ is fixed, there still is a degree of freedom for the determination of $q$, because $q$ depends on $F$ instead of $d\bar{F}^2/d\psi$, as can be seen in Eq. (\ref{q1}). In ECOM, this degree of freedom can be removed by specifying the value of either the poloidal current or the safety factor at a certain radial location.

There are several options for the scaling of the normalized equilibrium in ECOM, with corresponding namelist variable ISCALE. They are summarized in Table \ref{tab:iscale}. If ISCALE=0 or ISCALE=1, $\lambda$ is not rescaled, so that the total toroidal current is the normalized total toroidal current $I_{p}^{N}$. When ISCALE=0, the degree of freedom associated with $q$ is removed by specifying either $F(\Psi=\Psi_0)$ or $F(\Psi=\Psi_b)$. For $F(\Psi=\Psi_0)$, the additional namelist variable IFPOL needs to be set to 0, while for $F(\Psi=\Psi_b)$ IFPOL needs to be set to 1. When ISCALE=1, $q(\Psi=\Psi_0)$ is specified. If ISCALE=2 or ISCALE=3, $\lambda$ is rescaled so that the total toroidal current is adjusted to the desired toroidal current $I_{p}^{D}$. The choices for constraining the $q$ profile are the same as before: when ISCALE=2, either $F(\Psi=\Psi_0)$ or $F(\Psi=\Psi_b)$ is given, depending on the value of IFPOL, and when ISCALE=3, $q(\Psi=\Psi_0)$ is given. Finally, a last option to fix $\lambda$ in ECOM is to specify $q(\Psi=\Psi_0)$ and either $F(\Psi=\Psi_0)$ or $F(\Psi=\Psi_b)$. This option corresponds to ISCALE=4.
       
\begin {table*}
\caption {Options to scale the equilibrium in ECOM } 

\begin{center}
\begin{tabular}{ ||  c  || c || c ||}
\hline
Namelist &  Constraint to rescale $\lambda$& Constraint on $q$ profile\\
\hline
ISCALE=0  & None  &  $F(\Psi=\Psi_0)$ or $F(\Psi=\Psi_B)$ \\
\hline
ISCALE=1 &  None &  $q(\Psi=\Psi_0)$ \\
\hline
ISCALE=2 & $I_{p}$  & $F(\Psi=\Psi_0)$ or $F(\Psi=\Psi_B)$\\
\hline
ISCALE=3 & $I_{p}$  &  $q(\Psi=\Psi_0)$ \\
\hline
ISCALE=4 &  $F(\Psi=\Psi_0)$ or $F(\Psi=\Psi_B)$ and $q(\Psi=\Psi_0)$  & determined  \\
\hline
\end{tabular}
\end{center}

\label{tab:iscale}
\end {table*}

\subsection{Evaluation of the parallel current density}
ECOM includes the option to evaluate the neoclassical parallel current density using a reduced description for the bootstrap current, either based on the Hirshman model \cite{Hirshman} or on the Sauter model \cite{Sauter}, and the Sauter formula for the Ohmic current \cite{Sauter}. The namelist variables associated with these capabilities are IBSCUR and IJBSMODEL. IBSCUR must be set to 1 for ECOM to calculate the parallel current, and the model ECOM uses for the calculation of the bootstrap current depends on the value of the variable IJBSMODEL. If IJBSMODEL=1, the Hirshman model is used, if IJBSMODEL=2, the Sauter model is used. 

Consider the parallel current density in the formula
     \begin{eqnarray}
     \overline{J_\|R}(\psi)&\equiv&\frac{\langle \mathbf{J} \cdot \mathbf{B} \rangle }{\langle \mathbf{B} \cdot \nabla \phi \rangle}.
      \end{eqnarray}
The contribution of the ohmic current to the parallel current is determined by the loop voltage and the neoclassical resistivity according to 
     \begin{eqnarray}
      \overline{J_\|R}_{O}(\psi)&=& \sigma_{neo} \frac{V_{loop}}{2\pi},
        \end{eqnarray}
where $\sigma_{neo}$ is the neoclassical resistivity, and $V_{loop}$ is the loop voltage. In ECOM, the value of the loop voltage is specified by the namelist variable VLOOP0, in unit of volts, and  the Sauter model is implemented for the evaluation of the neoclassical resistivity $\sigma_{neo}$:
    \begin{eqnarray}
      \sigma_{neo} = \sigma_{spitz}\left[1-\left(1+\frac{0.36}{Z_i}\right)f_{teff(33)} +\frac{0.59}{Z_i}f_{teff(33)}^2-\frac{0.23}{Z_i}f_{teff(33)}^3\right],
        \end{eqnarray}
        where $ \sigma_{spitz}$ is the Spitzer resistivity as defined in \cite{Sauter}, $Z_i$ is the ion charge, and $f_{teff(33)}$ is given by Eq. (13b) in \cite{Sauter}. To evaluate $f_{teff(33)}$, the effective passing particle fraction $f_p(\psi)$ is calculated on each flux surface using the following formula
     \begin{eqnarray}
     f_p(\psi)=\frac{3}{4}\langle |B|(\psi,\theta)^2\rangle \int_0^{1/B_{max}(\psi)} dy \frac{ydy}{\langle\sqrt{1-y|B|(\psi,\theta)}\rangle} \label{fp},
          \end{eqnarray}
where $B_{max}$ is the maximum value of the magnetic field on the flux surface. The integral is computed numerically with a Chebyshev-Gauss quadrature.
   
If IJBSMODEL=1, ECOM evaluates the contribution of the bootstrap current $\overline{J_\|R}_{B}(\psi)$ to the parallel current with the Hirshman model \cite{Hirshman}. The quantity $\langle\mathbf{J}\cdot\mathbf{B}\rangle_{B}$ is calculated using the formulae (23) to (25) in Reference \cite{lutjens1996chease} and the passing particle fraction $f_p$ in Eq. (\ref{fp}) of the present article. If IJBSMODEL=2, ECOM relies on the Sauter model \cite{Sauter} to compute $\langle\mathbf{J} \cdot \mathbf{B} \rangle_B$. Specifically,
          \begin{eqnarray}
           \langle \mathbf{J} \cdot \mathbf{B} \rangle_B=-\frac{d p}{d\Psi}\frac{F}{1+\eta_i}\left[\mathcal{L}_{31}+\eta_i(\psi)(\mathcal{L}_{31}+0.5\mathcal{L}_{32}+0.5\mathcal{L}_{34}\alpha)\right)
           \end{eqnarray}
     where $\eta_i=(d \ln n/d\psi)^{-1}(d \ln T_i/d\psi)$, the electron and ion densities are the same, $n_e=n_i=n$, and the electron and ion temperatures are also assumed to be equal for simplicity, $T_e=T_i$. $\mathcal{L}_{31}$,$\mathcal{L}_{32}$, $\mathcal{L}_{34}$, and $\alpha$ are all defined in \cite{Sauter}, and are evaluated using $f_p$ in Eq. (\ref{fp}). 

\subsection{MHD stability}     
If the namelist variable ISTABILITY is set to 1, ECOM verifies whether the computed equilibrium crosses or not the Troyon limit \cite{Troyon}. This limit is thought to ensure the no-wall stability of the equilibrium to the $n=1$ internal kink mode, as well as ballooning, and external ballooning-kink modes, and can be expressed in two equivalent ways \cite{Freidberg_IdealMHD},  
   \begin{eqnarray}
   \beta_{T1}&=&0.028\frac{I_p}{aB_0}>\beta, \\
   \beta_{T2}&=&0.14\frac{a\kappa}{R_0q^{\star}}>\beta .  
  \end{eqnarray}
where $q^{\star}=2B_0A_0/(\mu_0 R_0 I_p)$, $A_0=\int_1^0 d\psi I_D(\psi)$ is the total poloidal cross section area, $I_D(\psi)=\int_0^{2\pi} d\theta ({J}/{R})$, and $\beta$ has been defined in Eq. (\ref{beta}).

In addition, ECOM also checks the Mercier criterion for stability against interchange modes \cite{Mercier} on each flux surface, given by
  \begin{equation}
 -D_I=\left(\frac{d p/d \Psi}{d q/d \Psi}F{I_G}\frac{|\lambda|^3}{2\pi}-\frac{1}{2}\right)^2+\frac{d p/d \Psi}{(dq/d \Psi)^2}\frac{\lambda^2}{4\pi^2}\left(\frac{d I_B}{d \Psi}-I_H\lambda^2\frac{d p}{d \Psi}\right)\left(F^2I_E\lambda^2+I_C\right)>0,
 \end{equation}
 where 
 \begin{equation}
 I_E(\psi)=\int_0^{2\pi} d\theta \frac{J }{|\nabla \psi|^2R^2},\qquad
  I_G(\psi)=\int_0^{2\pi} d\theta \frac{J}{|\nabla \psi|^2} ,\qquad 
   I_H(\psi)=\int_0^{2\pi} d\theta \frac{JR^2}{|\nabla \psi|^2}.
 \label{int456}
  \end{equation}

\subsection{Miller parametrization of the flux surfaces}

If the namelist variable IFITMIL is set to the value 1, ECOM uses a nonlinear least square method to compute the Miller parametrization that best fits a given flux surface of interest. The details of the fitting method can be found in Appendix \ref{sec:Miller}. The outputs of the calculation are the Miller parameters $\kappa$, $\delta_m$, $a/R_{m0}$, $dR_{m0}/d\rho$, $d\kappa/d\rho$, $d\delta_m/d\rho$, $q$, $\hat{s}$, and $\alpha_m$, which can for example be used in ballooning stability studies and in gyrokinetic codes. $\alpha_m$ is given by the expression \cite{miller}
 \begin{equation}
 \alpha_m=-\frac{1}{2\pi^2}\frac{dV}{d\Psi} \sqrt{\frac{V}{2\pi^2R_{m0}}}\mu_0\frac{dp}{d\Psi}.
   \end{equation}
   
\section{Accuracy and speed}\label{sec:Accuracy}
In this section, we consider two examples to compare the performance of ECOM with that of the popular G-S solver CHEASE \cite{lutjens1996chease}. The first example corresponds to a family of equilibria originally studied by Solov'ev \cite{solovev1968}, for which simple analytic expressions can be written for the solution $\Psi$. These equilibria are particularly advantageous for detailed error analysis, but lack generality in the sense that the G-S equation is linear and does not have to be solved as an eigenvalue problem. In Section \ref{sec:exnl} we thus consider a more general equilibrium, with $p$ and $F$ profiles chosen in such a way that the G-S equation is nonlinear and has to be solved as an eigenvalue problem. Every computational test in this article is conducted using a single core 2.6GHz AMD Opteron processor with 8GB of memory.  

\subsection{Example 1: Solov'ev profiles}\label{sec:Solov'ev}
For the first example, we consider the Solov'ev profiles $\mu_{0}{p}(\Psi)=-C_s\Psi$ and ${F}(\Psi)=F_B$, where $C_s$ and $F_B$ are constants. The G-S equation then reduces to $\Delta^{*}\Psi=C_sR^2$, and an up-down symmetric solution is given by the following expression \cite{lutjens1996chease}
\begin{eqnarray}
\Psi(R,Z)=\frac{\kappa F_B}{2R_0^3q_0} \left[\frac{1}{4}(R^2-R_0^2)^2+\frac{1}{\kappa^2}R^2Z^2-a^2R_0^2\right],\label{solo2}
\end{eqnarray}
where $C_s= F_B(\kappa+1/\kappa)/({R_0^3q_0})$, $R_0$ and $q_0$ are the major radius and the safety factor at the magnetic axis, and $a$ and $\kappa$ are the effective minor radius and elongation of the last closed flux surface, given by $\Psi=0$. A particularly convenient parametrization for the surface, which ECOM uses to compute the conformal map from the plasma boundary to the unit disk, is given by:
\begin{eqnarray}
R^2&=&R_0^2+2aR_0 \cos t \label{solob1},\\
Z&=&\kappa a\frac{R_0}{R} \sin t \label{solob2}.
\end{eqnarray}
The poloidal flux takes its minimum value $\Psi_{0}$ at the magnetic axis $R=R_0$, $Z=0$, with $\Psi_{0}$ given by 
\begin{eqnarray}
\Psi_0=-\frac{\kappa a^2}{2R_0 q_0},
\end{eqnarray}
The safety factor at a given flux surface $\Psi=\Psi_s$ can also be calculated exactly:
\begin{eqnarray}
q(\Psi=\Psi_s)&=& \frac{F}{\pi}\int_{R_{min}}^{R_{max}} \frac{dR\sqrt{1+(dZ/dR)^2}}{R\sqrt{(\partial\Psi/\partial R)^2+(\partial\Psi/\partial Z)^2}}\bigg |_{\Psi=\Psi_s}= \frac{F_B}{\pi}\int_{R_{min}}^{R_{max}} \frac{dR}{R|\partial \Psi/\partial Z|}\bigg|_{\Psi=\Psi_s}\\
&=& \frac{2q_0 R_0^3}{\pi}\int_{R_{min}}^{R_{max}} \frac{dR}{R^2\sqrt{(R^2-R_{min}^2)(R_{max}^2-R^2)}}\label{q3}\\
&=& \frac{2q_0}{\pi}\frac{R_0^3}{R_{min}^2R_{max}}E(k),\label{q4}
\end{eqnarray}
where $R_{max}$ and $R_{min}$ are the solutions of Eq. (\ref{solo2}) for $\Psi=\Psi_s$ and $Z=0$ satisfying $R_{max}>R_{min}>0$ and $R_{max}^2+R_{min}^2=2R_{0}^{2}$. Here, $E(k)$ is the complete elliptic integral of the second kind with modulus $k=\sqrt{1-(R_{min}/R_{max})^2}$. $E(0)=\pi/2$ and the integral formula 6 in Section $\mathbf{3.156}$ of Reference \cite{gradshteyn1965table} were used to derive Eq. (\ref{q4}). For simplicity, we define the magnetic shear $\hat{s}(\Psi)$ in terms of $\Psi$, giving the exact formula
\begin{eqnarray}
\hat{s}(\Psi)&=&\frac{\Psi}{q(\Psi)}\frac{dq(\Psi)}{d\Psi} =\frac{\Psi}{q(\Psi)}\frac{d}{d\Psi}\left[\frac{2q_0}{\pi}\frac{R_0^3}{R_{min}^2 R_{max}}E(k)\right]\\
&=&\frac{\Psi}{q(\Psi)}\frac{8q_0^2}{\pi\kappa F_B }\frac{R_0^6}{R_{min}^2 R_{max}^5}\frac{1}{k^4}\left[2\frac{1-k^2+k^4}{1-k^2}E(k)-(2-k^2)K(k)\right], \label{s0}
\end{eqnarray}
where $K(k)$ is the complete elliptic integral of the first kind. In ECOM, this Solov'ev case is computed when the namelist variables are chosen such that IPTYPE=0, IFTYPE=0, and IBTYPE=0.

\begin{figure*}
\includegraphics[scale=0.7]{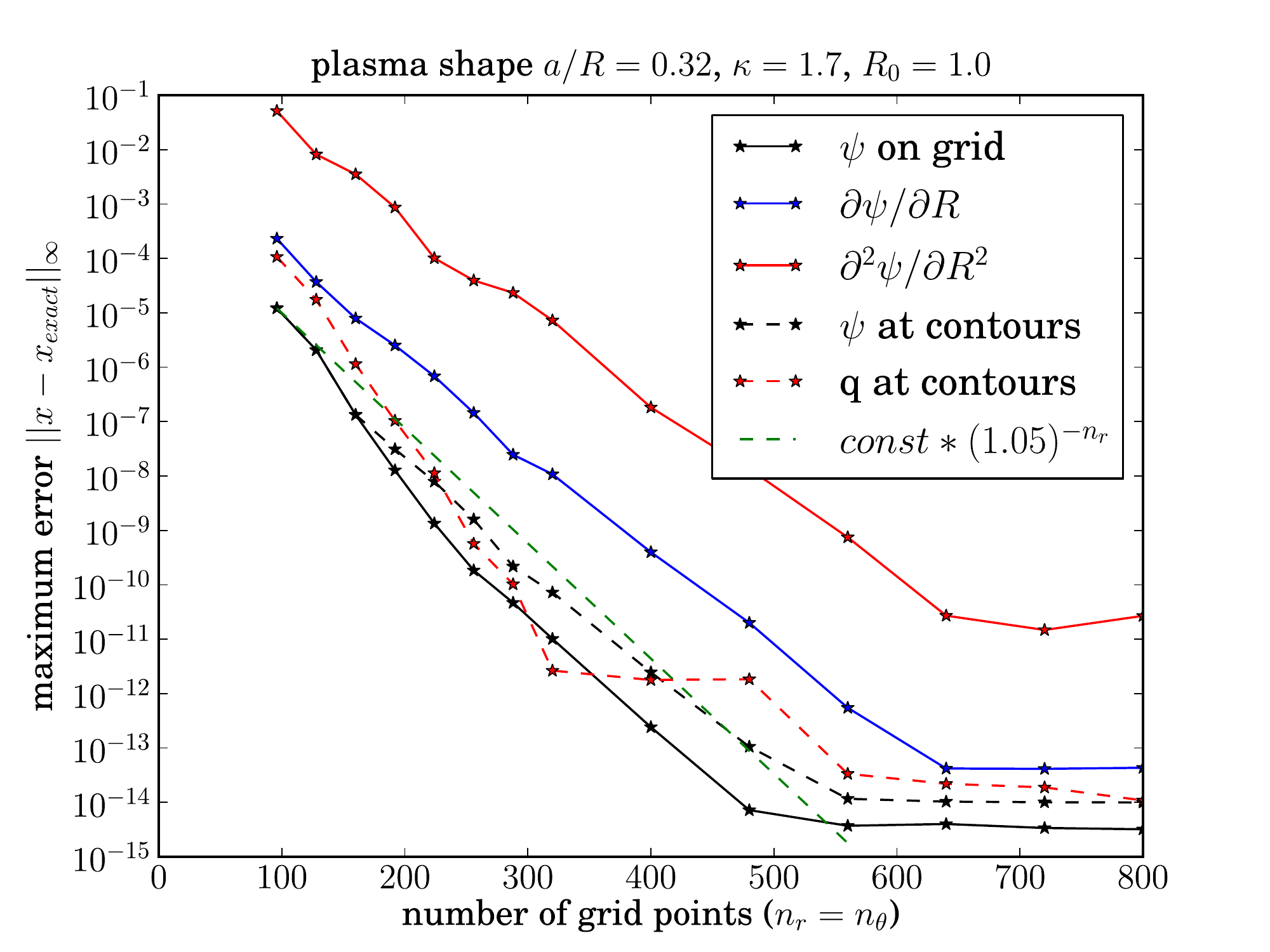}
\caption{Convergence of $\Psi$, its first and second radial derivatives, and the safety factor $q$ as a function of the number of grid points in one direction $N=n_{r}=n_{\vartheta}$. The exact equilibrium solution is given by Eq. (\ref{solo2}), and $R_0=1.0$, $a/R_0=0.32$, $\kappa=1.7$, $F_B=1.0$ and $q_0=1.0$. The number of flux contours constructed in postprocessing is $n_{f}=16$.} \label{Fig:maxerr}
\end{figure*}

Figure \ref{Fig:maxerr} shows the error in the $L^{\infty}$ norm between the numerical values of $\Psi$, $\partial\Psi/\partial R$, $\partial^{2}\Psi/\partial R^{2}$, and $q$ calculated with ECOM and the exact values computed from Eq. (\ref{solo2}) and Eq. (\ref{q4}), for the parameters $R_0=1.0$, $a/R_0=0.32$, $\kappa=1.7$, $F_B=1.0$ and $q_0=1.0$. The expression ``on grid" means that the error is evaluated at the $(R,Z)$ points of the grid on which ECOM solves the G-S equation. The curves labeled ``at contours", on the other hand, also include the error induced by the postprocessing steps described in Section \ref{sec:post}. Specifically, for all the flux contours constructed in the postprocessing phase, we compute the error between the value of $\Psi$ at the contour and the actual value of $\Psi$ at this location according to Eq. (\ref{solo2}). The largest of these errors over the whole domain $\Omega$ is used to plot the curve we call ``$\Psi$ at contours". The curve labeled ``$q$ at contours" represents the maximum error between $q$ at the contours as computed by ECOM according to Eq. (\ref{q1}) and the exact value of $q$ at these locations as given by Eq. (\ref{q4}). Figure \ref{Fig:maxerr} demonstrates the exponential convergence of the maximum error as the number of grid points is increased, as pointed out in the introduction. The green dashed line in Figure \ref{Fig:maxerr} indicates that the convergence rate is approximately $1.05^{-N}$. Figure \ref{Fig:maxerr} is also a proof that the derivatives of $\Psi$ have a convergence rate that is similar to that of $\Psi$. Note finally that the numerical methods used in the postprocessing steps to compute contours of constant flux lead to similar convergence rates and accuracy for $\Psi$ and $q$ on the contours.
   
\begin{figure*}
\includegraphics[scale=0.7]{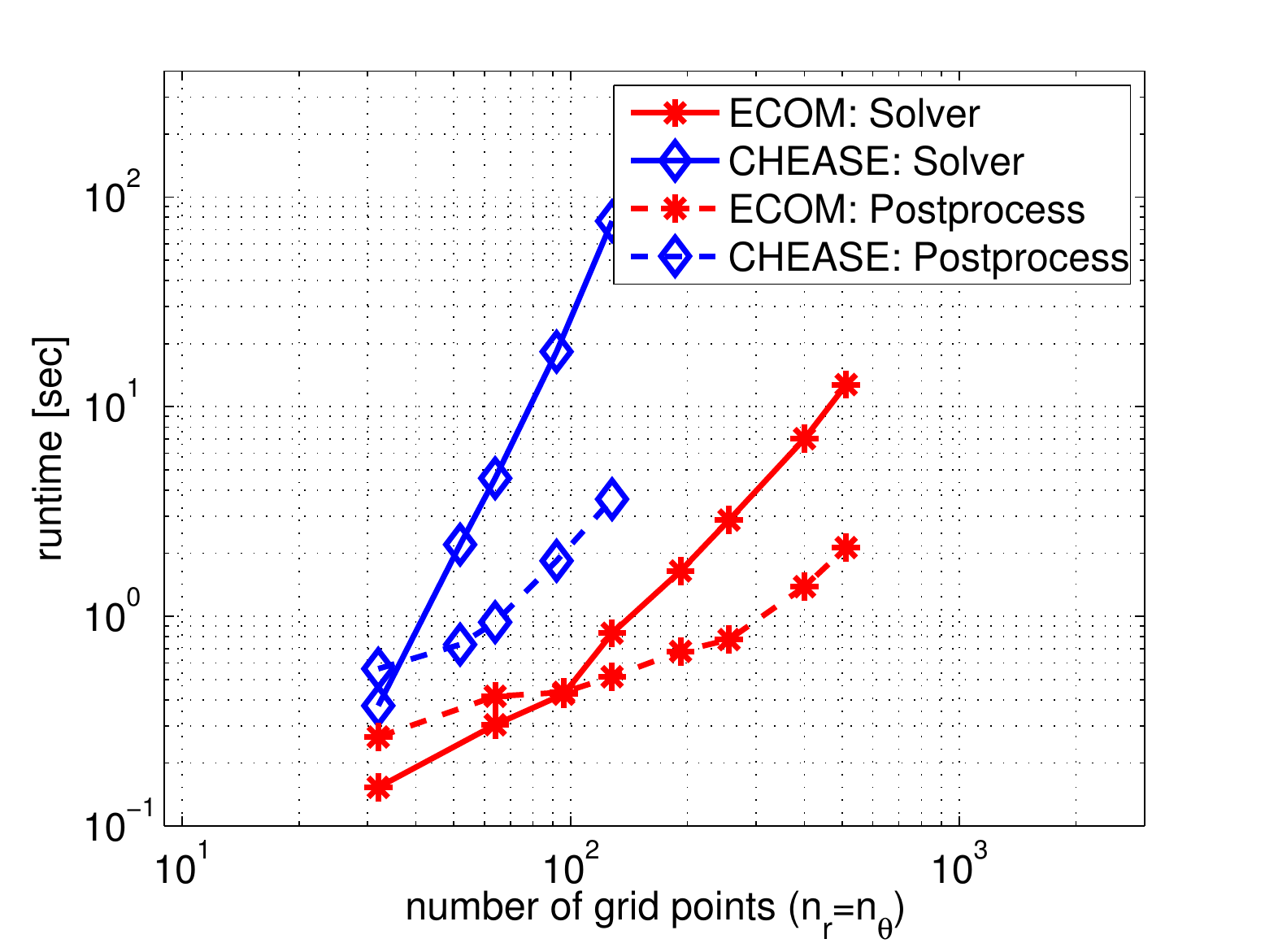}
\caption{Run time of the codes ECOM and CHEASE as a function of the number of grid points for the same Solov'ev equilibrium as Figure \ref{Fig:maxerr}. The solid lines represent the elapsed time until the converged $\Psi$ is obtained, and the dashed lines correspond to the postprocessing time for $n_f=16$ flux surfaces. For the data in this figure, each ECOM run computed the conformal map only once, with $n_1=n_\vartheta$ }\label{Fig:time}
\end{figure*}

At equal grid size, ECOM is much faster than CHEASE, as shown in Figure \ref{Fig:time}. The run time complexity of the solver is $O(n_rn_\vartheta \log n_\vartheta)$ instead of $O(n_r^{2}n_\vartheta^2)$ for typical finite element based codes solving the G-S equation. Note that the ``solver" part of ECOM represented by the solid line in Figure \ref{Fig:time} includes the run times of both the conformal mapping and the Poisson solver. The run time of the Poisson solver is the major contributor to the total run time in ECOM for typical grid sizes, because the Poisson solver is typically called 20 to 30 times during an equilibrium calculation while the conformal mapping routine is called at most a few times. The operation count of postprocessing depends on the number of contours $n_f$, which is typically smaller than the number of radial grid points: $n_f=10-30 < n_r$. The run time of postprocessing is relatively short because most computations are one dimensional, as described in Section \ref{sec:post2}.

\begin{figure*}
\includegraphics[scale=0.7]{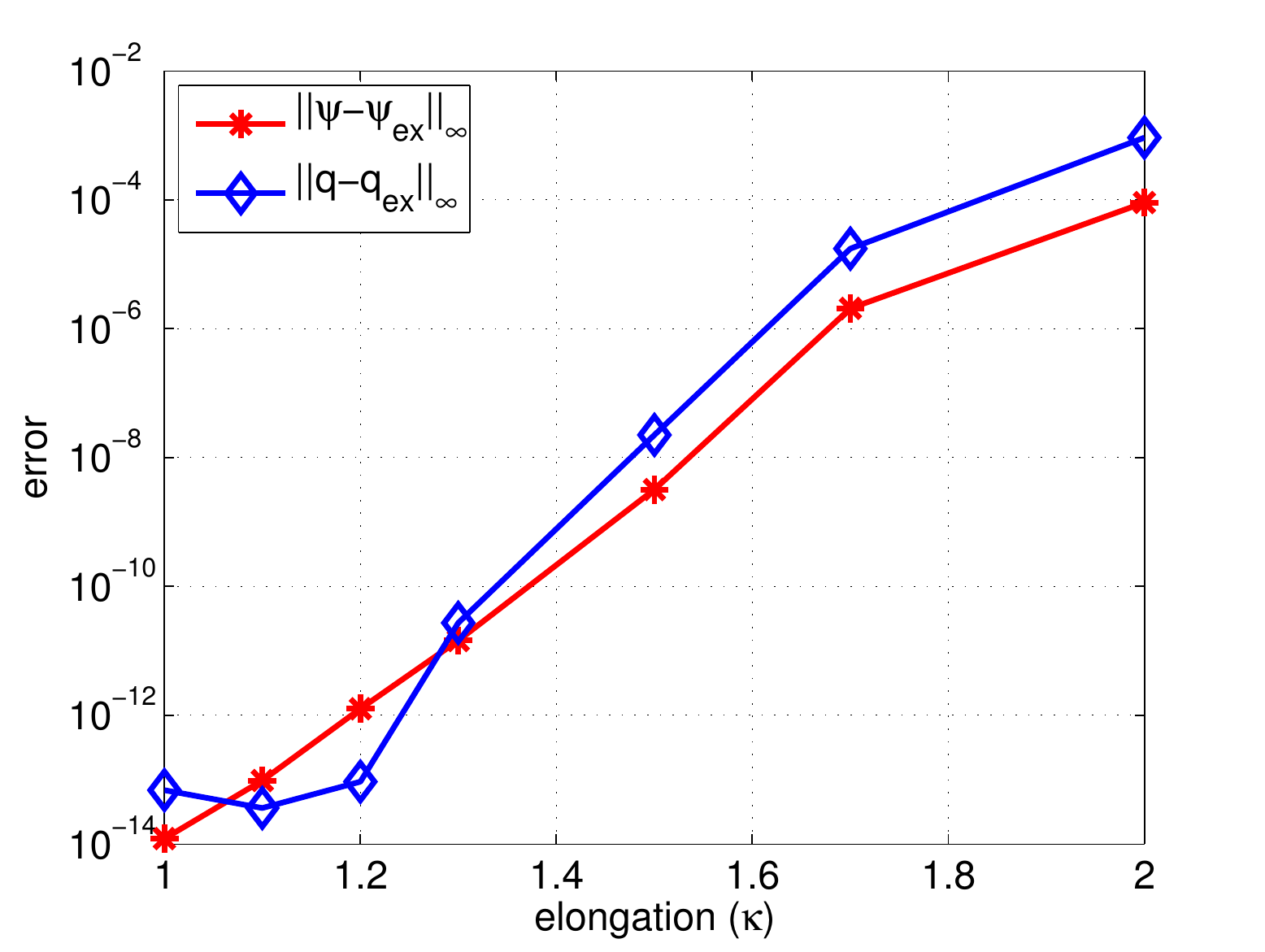}
\caption{Maximum error in $\Psi$ and in $q$ for the Solov'ev equilibrium studied in Section \ref{sec:Solov'ev} as a function of the elongation parameter $\kappa$. The error in $\Psi$ is measured on the $(R,Z)$ grid used to solve the G-S equation, while the error in $q$ is measured at the $n_{f}$ contours constructed during the postprocessing. Here, the grid resolution is $n_1=n_r=n_\theta=256$ and $n_f=16$}\label{Fig:elong}
\end{figure*}

Before comparing the accuracy of ECOM and CHEASE, it is instructive to look at the consequences of the grid crowding effect due to the conformal map on the accuracy of ECOM for shaped plasma equilibria. For fusion applications, a key question is how fast the numerical error evolves as the egg-shaped plasma cross section gets more and more elongated. Figure \ref{Fig:elong} provides an answer to that question for the Solov'ev equilibrium studied in this section, showing a significant deterioration of the accuracy of ECOM as the elongation of the last closed flux surface is increased. The consequence of this result is that for elongations corresponding to modern tokamaks and spherical tokamaks, $\kappa\simeq 1.5-2$, ECOM will often require a denser grid than FEM based G-S solvers to achieve the same accuracy. Since ECOM is much faster than these solvers in terms of work per grid point, and computes derivatives with high accuracy, ECOM remains very often more desirable that FEM based solvers, as we will show next. However, ECOM is not an attractive option to compute highly elongated equilibria, such as those in Field Reversed Configurations (FRCs) \cite{Freidberg_fu}, with $\kappa\sim10$.

\begin{figure*}
\includegraphics[scale=0.5]{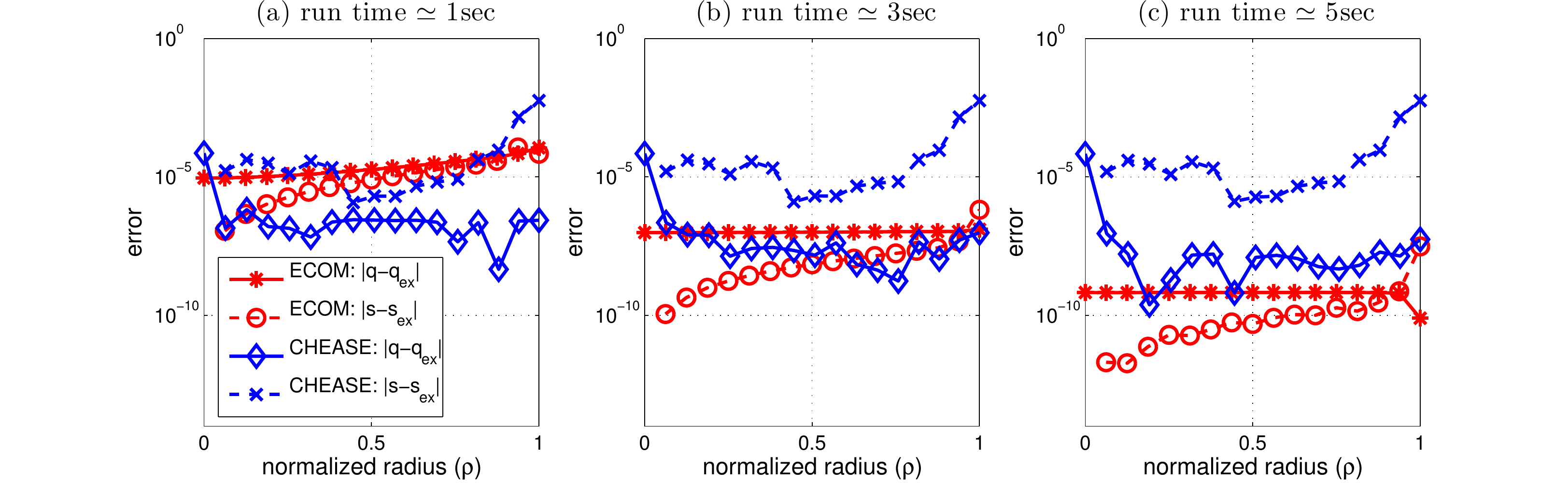}
\caption{Comparison of the error in ECOM and in CHEASE for the safety factor profile and the magnetic shear profile $\hat{s}=({\psi}/q)(dq/{d\psi})$ corresponding to the Solov'ev equilibrium studied in Section \ref{sec:Solov'ev} with $\kappa=1.7$ as a function of the normalized radius ${\rho}=\sqrt{(\Psi-\Psi_0)/(\Psi_b-\Psi_0)}$. Panel (a) was obtained for grids leading to a total run time of 1 second, panel (b) to a total run time of 3 seconds, and panel (c) to a total run time of 5 seconds. Here, $n_1=n_r=n_\vartheta$ and $n_f=16$}\label{Fig:qs}
\end{figure*}

We have just shown that ECOM is much faster than CHEASE for a given grid size, but that for elongated plasma shapes ECOM may need a denser mesh to achieve a desired accuracy, due to crowding effects. In this context, a fair comparison of the performance of the two codes is done by evaluating the accuracy of each solver for a given run time on the same machine. This is precisely the purpose of Figure \ref{Fig:qs}, which shows the numerical error in the safety factor $q$ and the magnetic shear $\hat{s}$ as a function of the normalized radius ${\rho}=\sqrt{(\Psi-\Psi_0)/(\Psi_b-\Psi_0)}$ for the Solov'ev equilibrium considered in this section and $\kappa=1.7$. We have chosen to focus on $q$ and $\hat{s}$ because MHD stability and turbulent transport are known to depend sensitively on these quantities. We compare ECOM (red markers) and CHEASE (blue markers) for three different run times. In Figure \ref{Fig:qs}-(a), the run time is 1 second, corresponding to a grid resolution of $n_r=32$ in CHEASE and $n_r=96$ in ECOM; in Figure \ref{Fig:qs}-(b), the run time is 3 seconds, corresponding to a grid size of $n_r=52$ in CHEASE and $n_r=192$ in ECOM; in Figure \ref{Fig:qs}-(c), the run time is 5 seconds for a grid resolution of $n_r=64$ in CHEASE and $n_r=256$ in ECOM\color{black}. One can see that for small grids, CHEASE computes the safety factor with a better accuracy than ECOM, a direct consequence of the crowding effect. However, even if in that case CHEASE calculates $q$ more accurately, the accuracy for the magnetic shear are comparable in ECOM and CHEASE. One reason for this is that in ECOM, we constructed the $n_{f}$ flux contours so that they would coincide with a global Chebyshev grid of size $n_{f}$ on the interval $\psi=0$ and $\psi=1$. ECOM can thus use spectral differentiation to compute $\hat{s}$, leading to a more limited loss of accuracy between $q$ and $\hat{s}$. The construction of such a Chebyshev grid for the flux variable is particularly convenient in ECOM because the piecewise Chebyshev grid used by the Poisson solver to discretize the radial direction in $D_{1}$ is well refined near the end points $\psi=0$ and $\psi=1$ of the interval. Radial derivatives of flux functions (e.g. $\hat{s}$, $dV/d\rho$ and $d\psi/d\rho$) are therefore calculated without significant loss of accuracy in ECOM.

As the grid size and computation time are increased, ECOM outperforms CHEASE, which is a direct result of the geometric convergence demonstrated in Figure \ref{Fig:maxerr}. For a run time of 3 seconds, $q$ is computed with similar accuracy in ECOM and CHEASE, but the error on $\hat{s}$ is more than 100 times smaller in ECOM. For a run time of 5 seconds, the error on both $q$ and $\hat{s}$ with ECOM is orders of magnitude smaller than the error obtained with CHEASE.

Although memory aspects rarely lead to severe constraints on G-S solvers in fusion applications, it is interesting to note that memory requirements are much smaller in ECOM than they are in CHEASE. As an empirical illustration of this, we observed that because of the memory limitations of the computer we used for our comparison studies, we were limited to grids smaller than $n_rn_\vartheta \lesssim 2\times10^{4}$ in CHEASE, whereas we computed equilibria in ECOM with grids $n_rn_\vartheta \geq10^{6}$ without any difficulty.

\subsection{Example 2: Nonlinear Grad-Shafranov equation}\label{sec:exnl}

For the second example, we choose pressure and current profiles in such a way that the G-S equation is an eigenvalue partial differential equation given by Eq. (\ref{eigen}). Specifically, we set the namelist variables IPTYPE and IFTYPE to 1, and specify the profile constants according to $p_{in}=2$, $p_{out}=1$, $F_{0\psi}=-1$, $F_{in}=2$, and $F_{out}=1$. In the three equilibria we study in this section, we will vary $p_{0\psi}$: for the first equilibrium, we set $p_{0\psi}=-0.01/\mu_{0}$, leading to a very small Shafranov shift, and we set $p_{0\psi}=-1/\mu_{0}$ for the next two equilibria, which consequently have a much larger Shafranov shift. To describe the last closed flux surface of the equilibrium, we set the namelist varibale IBTYPE to 1, corresponding to the Miller parametrization:
\begin{eqnarray}
R&=&R_{m0}+a \cos (t+\sin^{-1}\delta_m \sin t)  \label{millerb1},\\
Z&=&Z_{m0}+a\kappa \sin t \label{millerb2},
\end{eqnarray}
where the parameter $t$ goes from 0 to $2\pi$, $\kappa$ is the elongation and $\delta_m$ is the triangularity. For the purpose of comparison, we specify the same profiles and parametrization of the plasma boundary in the input file of CHEASE, and for both codes we look at the convergence of the poloidal magnetic field energy $W_p$, given by Eq. (\ref{polE}). This global, 0-D quantity has often been used to measure the convergence properties of CHEASE \cite{lutjens1992}. For the equilibrium under consideration, the analytic expression for $\psi$ is not known, and there does not exist a formula for $W_p$. We thus use the value of $W_{p}(n_{ref})$ calculated with a large number of grid points as the reference point for the convergence studies.

\begin{figure*}
\includegraphics[scale=0.6]{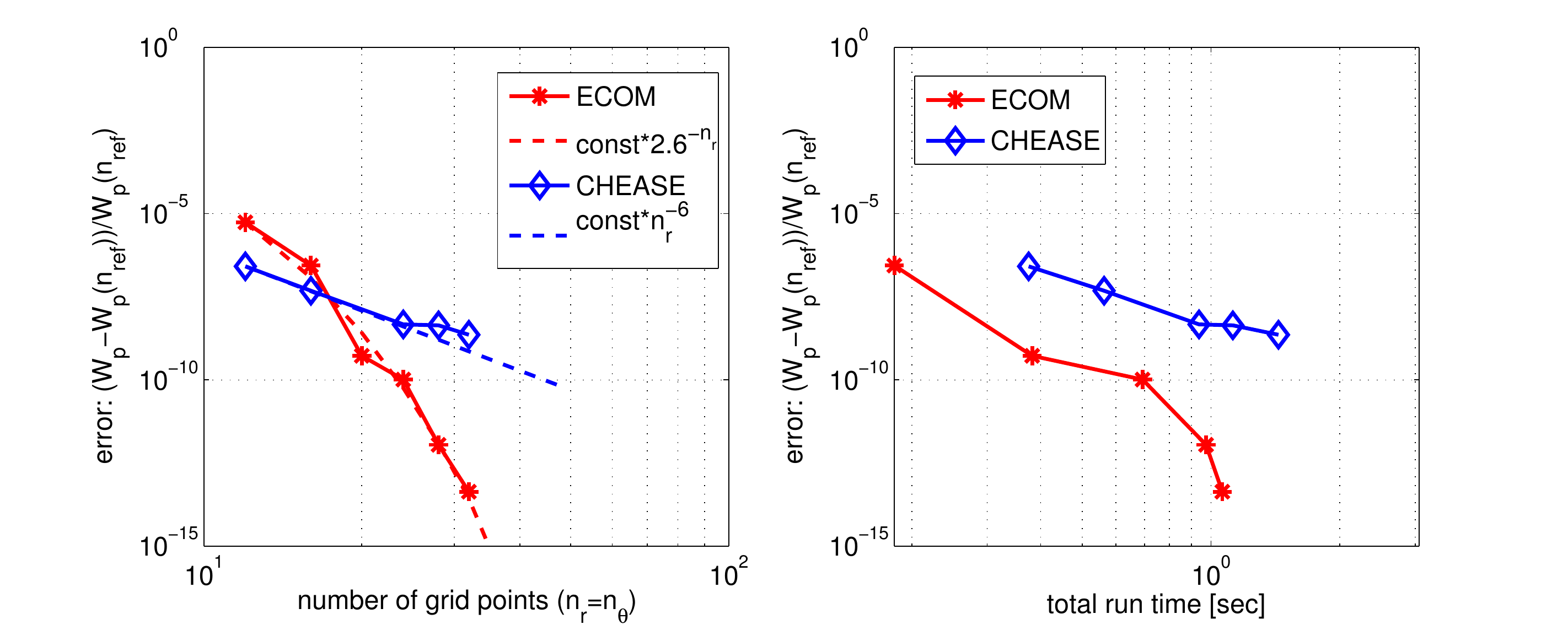}
\caption{Relative error in the poloidal magnetic energy $W_p$ for a circular tokamak equilibrium as a function of (a) the number of radial grid points ($n_r=n_\vartheta$) and (b) the total run time. $a/R_{m0}=0.32$, $\kappa=1.0$, $\delta_m=0.0$ and the pressure and current profiles are $\mu_0 d\bar{p}/d \psi=-0.01 (1-(1-\psi)^2)$ and $1/2(d \bar{F}^2/d \psi)=- (1-(1-\psi)^2)$, respectively, and ISCALE=0. For both CHEASE and ECOM, the reference values $W_{p}(n_{ref})$ was computed with a grid size $N=n_{ref}=48$, and the difference between the reference value of $W_{pE}(n_{ref})$ of ECOM and $W_{pC}(n_{ref})$ of CHEASE is $|W_{pE}(n_{ref})-W_{pC}(n_{ref})|/W_{pC}(n_{ref})=2.6\times10^{-9}$}\label{Fig:nl_kappa1_delta0}
\end{figure*}

\begin{figure*}
\includegraphics[scale=0.6]{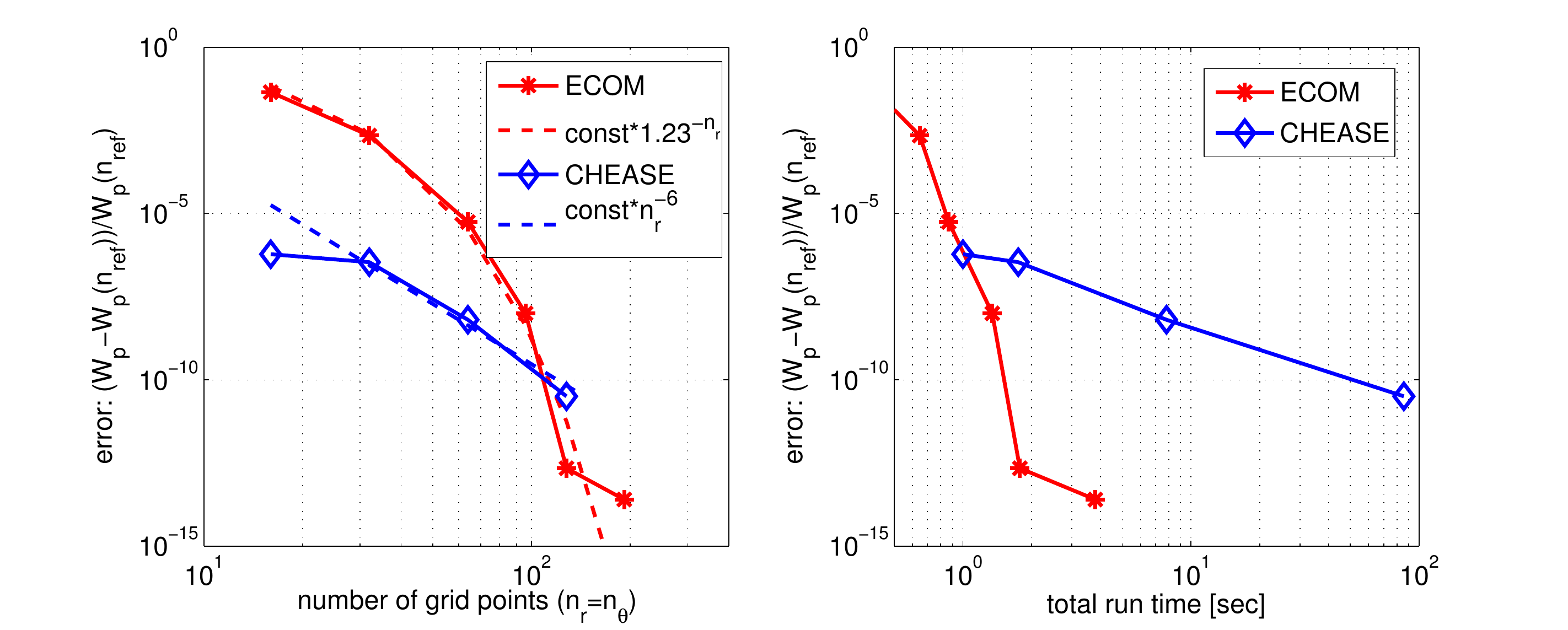}
\caption{Relative error in the poloidal magnetic energy $W_p$ for a tokamak equilibrium with triangularity as a function of (a) the number of grid points ($N=n_r=n_\vartheta$) and (b) the total run time. $a/R_{m0}=0.32$, $\kappa=1.0$, $\delta_m=0.33$, and the pressure and current profiles are $\mu_0d \bar{p}/d \psi=-(1-(1-\psi)^2)$ and $1/2(d \bar{F}^2/d \psi)=- (1-(1-\psi)^2)$ respectively, and ISCALE=0. $W_{pC}(n_{ref})$ was computed with a grid size $N=n_{ref}=136$, $W_{pE}(n_{ref})$ with a grid size $N=n_{ref}=256$, and $|W_{pE}(n_{ref})-W_{pC}(n_{ref})|/W_{pC}(n_{ref})=1.3\times10^{-9}$. Each ECOM run computed the conformal mapping twice.}\label{Fig:nl_kappa1}
\end{figure*}

\begin{figure*}
\includegraphics[scale=0.6]{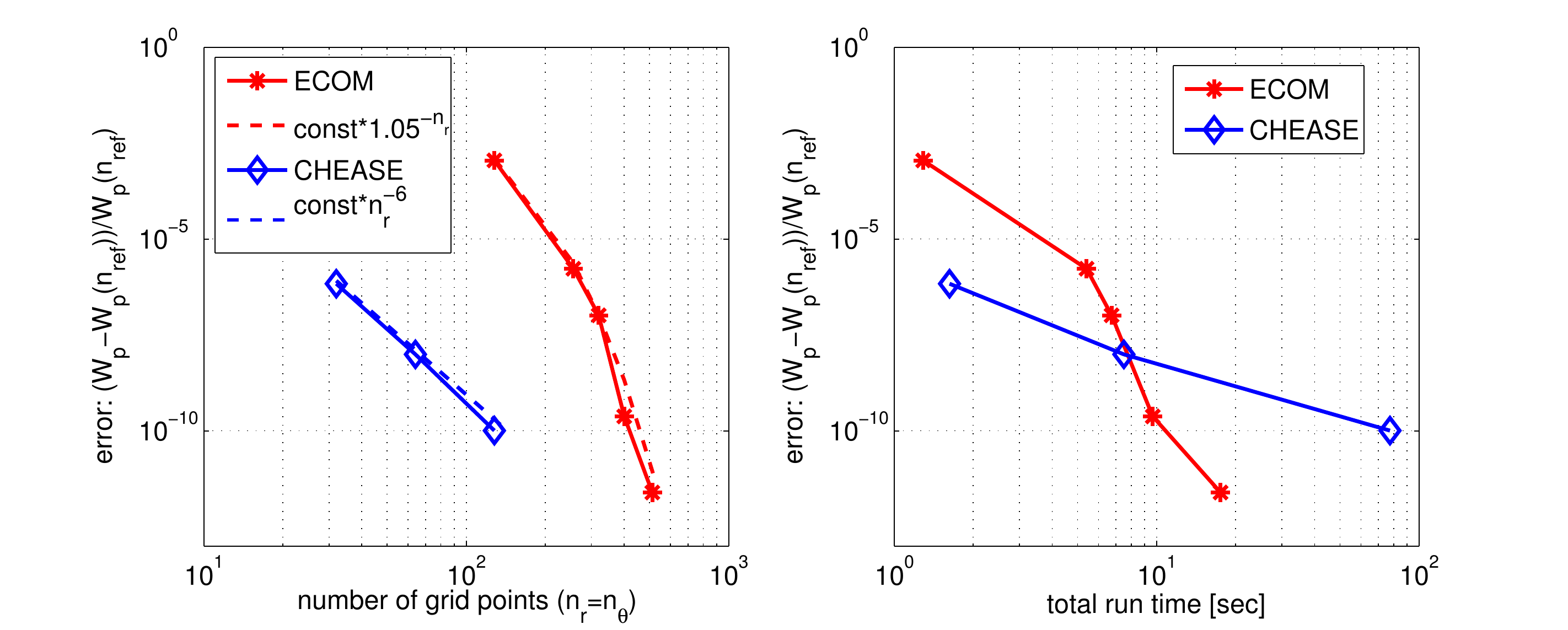}
\caption{Relative error in the poloidal magnetic energy $W_p$ for a tokamak equilibrium with triangularity and elongation as a function of (a) the number of grid points ($N=n_r=n_\vartheta$) and (b) the total run time. $a/R_{m0}=0.32$, $\kappa=1.7$, $\delta_m=0.33$ and the pressure and current profiles are $\mu_0 d \bar{p}/d \psi=-(1-(1-\psi)^2)$ and $1/2(d \bar{F}^2/d \psi)=- (1-(1-\psi)^2)$ respectively, and ISCALE=0. $W_{pC}(n_{ref})$ was computed with a grid size $N=n_{ref}=136$, $W_{pE}(n_{ref})$ with a grid size $N=n_{ref}=528$, and $|W_{pE}(n_{ref})-W_{pC}(n_{ref})|/W_{pC}(n_{ref})=7.6\times10^{-10}$. Each ECOM run computed the conformal map twice.}\label{Fig:nl_kappa1_7}
\end{figure*}

Figures \ref{Fig:nl_kappa1_delta0} -- \ref{Fig:nl_kappa1_7} show the convergence of $W_p$ in CHEASE and in ECOM as a function of the number of grid points and as a function of run time, for three different plasma shapes. Figure \ref{Fig:nl_kappa1_delta0} corresponds to a circular tokamak equilibrium with a small Shafranov shift. Since the domain $\Omega$ is a disk, ECOM computes such equilibria without conformal mapping, and this case allows us to focus on the error that is not due to the conformal mapping part of the solver. We observe that for a very small number of grid points, $N\leq 16$, CHEASE computes $W_{p}$ with higher accuracy than ECOM. This is somewhat artificial in that CHEASE uses a grid that is refined near the magnetic axis \cite{lutjens1996chease}, whereas in the absence of conformal mapping the center of the $(r,\theta)$ grid in ECOM does not coincide with the magnetic axis. The convergence rate of the poloidal magnetic energy in CHEASE is $N^{-6}$ as found in \cite{lutjens1992}, while the convergence rate in ECOM is found to be $2.6^{-N}$. As a result, for grids with $N> 16$ ECOM quickly becomes much more accurate than CHEASE. Finally, if we fix the run time instead of the grid size, we find that ECOM is always more accurate than CHEASE, as shown in Figure \ref{Fig:nl_kappa1_delta0}-(b).

In Figure \ref{Fig:nl_kappa1}, we consider an equilibrium with a larger Shafranov shift, a significant triangularity but with no elongation. Comparing Figure \ref{Fig:nl_kappa1_delta0} and Figure \ref{Fig:nl_kappa1}, we can see that the crowding effect inherent to conformal mapping results in a strong loss of accuracy, with a relative error in $W_{p}$ which is up to $10^4$ times larger at low grid resolutions. It also results in a reduction of the convergence rate from $2.6^{-N}$ to $1.23^{-N}$. Even if so, ECOM remains more accurate than CHEASE at fixed run time, as shown in Figure \ref{Fig:nl_kappa1}-(b). Figure \ref{Fig:nl_kappa1_7} corresponds to an ITER-like equilibrium \cite{Aymar}, with the same triangularity and pressure profile as Figure \ref{Fig:nl_kappa1}, but with elongation $\kappa=1.7$. As we would expect from Figure \ref{Fig:elong},  elongation amplifies the crowding effect, leading to further degradation of the accuracy. The convergence rate is reduced from $1.23^{-N}$ to $1.05^{-N}$. The strong dependency of the convergence rate on the plasma geometry and on crowding is an undesirable aspect of ECOM. In contrast, Figures \ref{Fig:nl_kappa1_delta0}-\ref{Fig:nl_kappa1_7} show that convergence in CHEASE is fairly insensitive of the shape of the plasma boundary. Despite this weakness, ECOM computes $W_{p}$ more accurately than CHEASE for run times longer than 8 seconds. Furthermore, the results in Section \ref{sec:Solov'ev} suggest that the run time threshold is lower for local quantities, in particular if these quantities depend on derivatives of flux functions, such as $\hat{s}$ or high order derivatives of the $\psi$, such as the local magnetic shear.
  
\section{Equilibria with toroidal flows}\label{sec:torflow}

Large equilibrium flows are observed in tokamak experiments \cite{Scott,Lao_shear,Menard}, and flows and flow shear are thought to have a strong influence on the stability and transport properties of the plasma \cite{Cooper,Hameiri,MillerBalloon,Waelbroeck,Lao_shear,Furukawa2001,Menard,Parra_PRL,Barnes_PRL,Highcock}. When the flow speed is of the same order as the sound speed, the inertial term in the pressure balance relation can no longer be ignored, and MHD equilibrium force balance is given by
\begin{equation}
\rho\mathbf{u} \cdot \nabla \mathbf{u} +\nabla p =\mathbf{J} \times \mathbf{B}  \label{EQflow1},
\end{equation}
where $\rho=m_{i}n$, $m_{i}$ is the ion mass, $n$ the ion density, and $\mathbf{u}$ the plasma flow. The plasma flow $\mathbf{u}$ in Eq. (\ref{EQflow1}) must also satisfy the steady-state version of Faraday's law in the ideal MHD model \cite{goedbloed2010advanced}:
\begin{equation}\label{Faraday}
\nabla\times(\mathbf{u}\times\mathbf{B})=\mathbf{0}.
\end{equation}
Poloidal flows are damped by neoclassical viscosity and expected to be much smaller than the ion sound speed in toroidally axisymmetric equilibria, except perhaps near the edge \cite{Hinton,hassam1,hassam2}. It is therefore a good approximation to only retain the effect of toroidal flows in Eq. (\ref{EQflow1}). The most general toroidal flow $\mathbf{u}=u\mathbf{e}_{\phi}$ satisfying Eq. (\ref{Faraday}) can be written as $u_{\phi}=R\Omega_{\phi}(\Psi)$. For a purely toroidal flow, MHD force balance thus takes the form
\begin{equation}\label{eq:momentum_general}
-\rho R\Omega_{\phi}^{2}(\Psi)\mathbf{e}_{R}=\mathbf{J}\times\mathbf{B}-\nabla p,
\end{equation}
where $\mathbf{e}_{R}=\nabla R$. In a general axisymmetric geometry, $p$ only depends on two variables. Since we know that for static equilibria $p$ is a function of $\Psi$ only, we choose $R$ and $\Psi$ as the two independent variables for the pressure profile in axisymmetric equilibria. We then have $\nabla p=\partial p/\partial R \nabla R+\partial p/\partial\Psi\nabla\Psi$, and dotting Eq. (\ref{eq:momentum_general}) with $\mathbf{B}$ yields an equation for the $R$ dependence of the pressure:
\begin{equation}\label{eq:first_relation}
\rho R\Omega_{\phi}^{2}(\Psi)=\frac{\partial p(R,\Psi)}{\partial R},
\end{equation}
where we have used the fact that $\mathbf{B}\cdot\nabla\Psi=0$. When Eq. (\ref{eq:first_relation}) is satisfied, Eq. (\ref{eq:momentum_general}) can be written as
\begin{equation}\label{eq:momentum_psi}
\mathbf{J}\times\mathbf{B}=\frac{\partial p (R,\Psi)}{\partial \Psi}\nabla\Psi.
\end{equation}
By dotting this equation with $\mathbf{J}$, it is easy to show that $RB_{\phi}=F(\Psi)$ as in the static case, and Equation (\ref{eq:momentum_psi}) becomes the following modified G-S equation for the flux function $\Psi$ in the presence of a toroidal flow: 
\begin{equation}\label{eq:GS_axi_flow_final}
\Delta^{*}\Psi=-\mu_{0}R^{2}\frac{\partial p(R,\Psi)}{\partial\Psi}-\frac{1}{2}\frac{dF^{2}}{d\Psi}
\end{equation}
There are two well known situations for which Eq. (\ref{eq:first_relation}) can be integrated analytically. The first situation corresponds to the assumption that the entropy $S\equiv p\rho^{-\gamma}$ is only a function of the poloidal flux \cite{goedbloed2010advanced}, the second situation corresponds to the assumption that the temperature is a flux function because of the high thermal conductivity along the magnetic field lines in fusion grade plasmas \cite{Maschke1980,furukawa2000tokamak,jardin2010computational}. Currently, ECOM only treats the latter case. Neglecting temperature anisotropy, we write $p(R,\Psi)=2n(R,\Psi)T(\Psi)$, with $T(\Psi)=0.5[T_{i}(\Psi)+T_{e}(\Psi)]$ a species averaged temperature, and integrate Eq. (\ref{eq:first_relation}) to find \cite{jardin2010computational}:
\begin{equation}
p(R,\Psi)=p_0(\Psi)\exp\left[\frac{p_\Omega(\Psi)}{p_0(\Psi)}\left(\frac{R^2}{R_{0}^2}-1\right)\right], \label{Pflow}
\end{equation}
where $p_0(\Psi)=p(R_{0},\Psi)$ and $p_\Omega(\Psi)=n_0(R_{0},\Psi)m_i\Omega_\phi(\Psi)^2R_{0}^2/2$ is the kinetic pressure due to the ion toroidal flow. ECOM uses the numerical scheme described in Section \ref{sec:GS} to solve the normalized version of Eq. (\ref{eq:GS_axi_flow_final}), 
\begin{equation}\label{eq:GS_axi_flow_normalized}
\Delta^{*}\psi=-\lambda\left(\mu_{0}R^{2}\frac{\partial \bar{p}(R,\psi)}{\partial\psi}-\frac{1}{2}\frac{d\bar{F}^{2}}{d\psi}\right)
\end{equation}
with the normalized pressure term given by
\begin{equation}
\bar{p}(R,\psi)=\bar{p}_0(\psi)\exp\left[\frac{\bar{p}_\Omega(\psi)}{\bar{p}_0(\psi)}\left(\frac{R^2}{R_{0}^2}-1\right)\right],\;\;\frac{d\bar{p}_{0}}{d\psi}=\frac{dp_{0}}{d\Psi}\label{Pflownormalize}
\end{equation}
In ECOM, the profile $\bar{p}_{0}(\psi)$ is specified in the same way as $\bar{p}(\psi)$ is in the static case, with the same namelist variable IPTYPE, and the same options. If IPTYPE=1 or IPTYPE=2 and IPTABLE=0, integration is required to obtain $\bar{p}_{0}(\psi)$ from its flux derivative. ECOM uses Chebyshev-Gauss quadrature on the global Chebyshev grid for $\psi$ to compute these integrals. There are several options to specify the kinetic pressure profile $\bar{p}_\Omega(\psi)$, with corresponding namelist parameter ITFTYPE. If ITFTYPE=1, $\bar{p}_\Omega$ is such that the toroidal Mach number $M=\sqrt{2\bar{p}_\Omega/\bar{p}_0}$ has the same value at all radii. If ITFTYPE=2, $\bar{p}_{\Omega}$ is given by the explicit formula $\bar{p}_\Omega(\psi)=\bar{p}_{\Omega0}(1-(1-\psi)^{p_{\Omega in}})^{p_{\Omega out}}$. If ITFTYPE=3, $\bar{p}_\Omega$ is given as a numerical table in terms of $\psi$, as is also done for the pressure and poloidal current profiles.

\begin{figure*}
\includegraphics[scale=0.6]{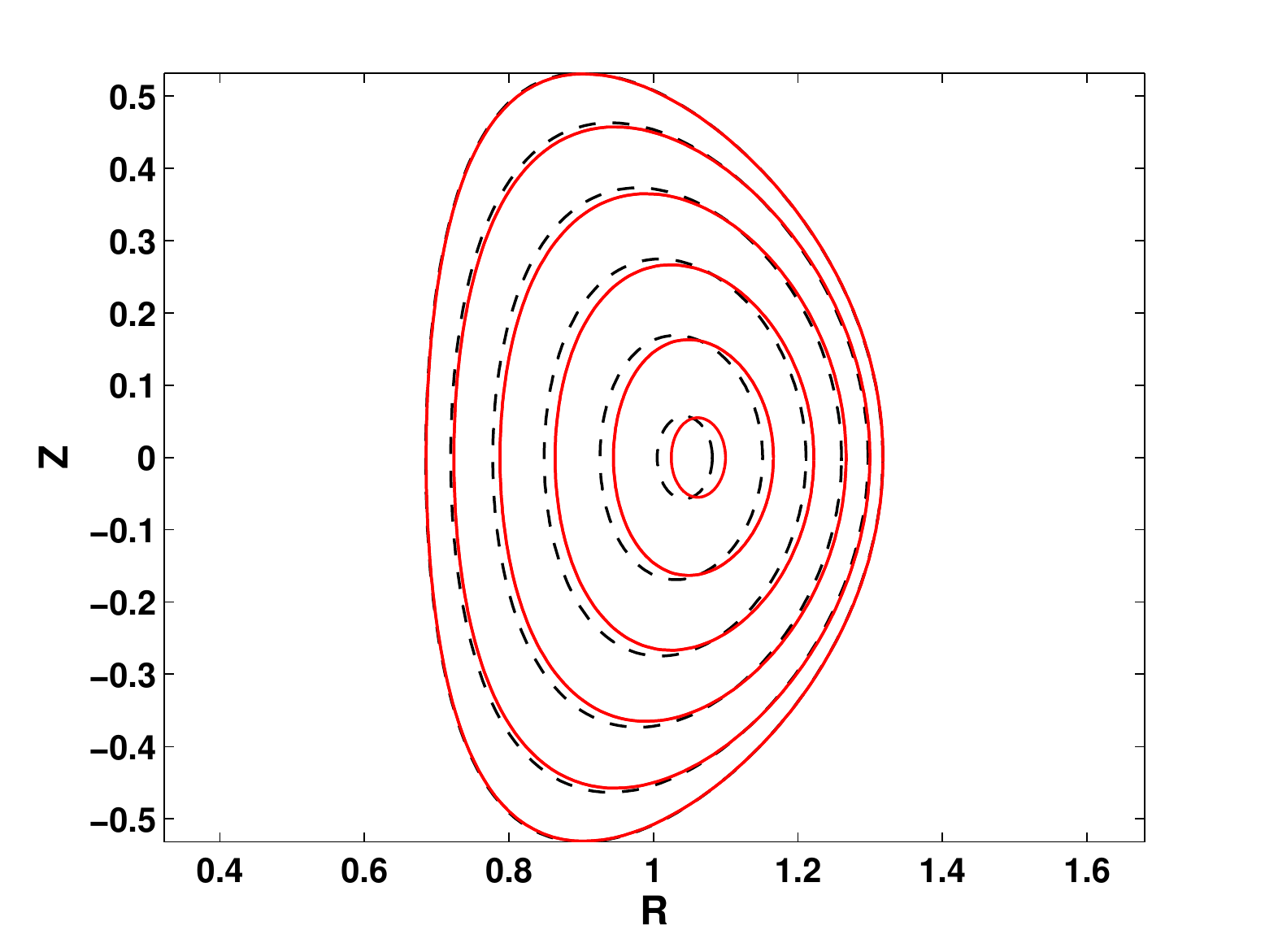}
 \caption{Flux contours for an equilibrium without toroidal flow (black dashed line, M=0.0) and with toroidal flow (solid red line, $M=1.0$). The shape of the boundary is given by the Miller parametrization with $\kappa=1.7$, $a/R_{M0}=0.32$, $\delta=0.33$, and the normalized pressure and poloidal current profiles are $\mu_0{d\bar{p}}/{d \psi}=1-(1-\psi)^2$ and $(1/2)({d\bar{F}^{2}}/{d \psi})=0.1(1-(1-\psi)^2)$, giving $I_{p}=1.0$ [MA] and $\beta_P=0.81$ }\label{Fig:contour}
\end{figure*}

Among the expressions ECOM uses in postprocessing to evaluate flux functions and figures of merit, only a few need to be modified in the presence of an equilibrium toroidal flow. Equation (\ref{GSrsec3}) becomes
\begin{equation}
 \frac{d I_A}{d \psi}=-\lambda\left[\mu_{0} \frac{d\bar{p}_0}{d\psi}I_{B1}(\psi)+\mu_{0}\frac{d\bar{p}_\Delta}{d\psi}I_{B2}(\psi) +\frac{1}{2}\frac{d {\bar {F}^2}}{d \psi} I_C(\psi)\right]\label{GSrsec3F}.
\end{equation}
where  $d \bar{p}_{\Delta}/d \psi=d \bar{p}_\Omega/d \psi-(\bar{p}_\Omega/\bar{p}_0)d \bar{p}_0/d \psi$, and $I_{B1}(\psi)$ and $I_{B2}(\psi)$ are defined by
\begin{equation}
  I_{B1}(\psi)=\int_0^{2\pi} d\theta J\exp\left[\frac{\bar{p}_\Omega(\psi)}{\bar{p}_0(\psi)}\left(\frac{R^2}{R_{0}^2}-1\right)\right]\qquad,\qquad 
    I_{B2}(\psi)=\int_0^{2\pi} d\theta J\left(\frac{R^2}{R_{0}^2}-1\right)\exp\left[\frac{\bar{p}_\Omega(\psi)}{\bar{p}_0(\psi)}\left(\frac{R^2}{R_{0}^2}-1\right)\right]\label{int2F}
\end{equation}
As a result, the intermediate step we use to derive Eq. (\ref{torcur1}) takes a slightly different form, but Eq. (\ref{torcur1}) itself does not change: the relation $I_{\phi}(\psi)=I_{A}(\psi)/(\lambda\mu_{0})$ still holds. Finally, the expression for the volume averaged pressure is now given by
\begin{equation}
\langle p\rangle_{V}=2\pi\frac{\int_{1}^{0}d\psi I_{B1}(\psi)\bar{p}_{0}(\psi)}{\lambda V_{0}}
\label{pres_flow}
\end{equation}
Figure \ref{Fig:contour} shows the flux contours of a stationary equilibrium with toroidal flow computed with ECOM and the flux contours of the corresponding static equilibrium also computed with ECOM. For that example, we chose $p_{\Omega}$ so that the toroidal Mach number is uniform with value 1. The flux contours of the stationary equilibrium are the red lines, and the flux contours of the static equilibrium are the black dashed lines. We can clearly see the expected outward shift of the magnetix axis \cite{green1973effect,furukawa2000tokamak}.  

\section{Discussion}\label{sec:dis}

ECOM uses conformal mapping from the plasma domain to the unit disk to decouple the numerical issues associated with the plasma geometry from the rest of the problem. Once on the unit disk, ECOM relies on fairly standard high order methods based on the FFT for the angular dependence and a Green's functions formulation for the radial dependence to solve the mapped partial differential equation describing the plasma equilibrium. This approach is not only conceptually elegant, it is also effective. By studying a static equilibrium with Solov'ev profiles, we showed in this article that the solution of the Grad-Shafranov equation as well as its first and second derivatives converge exponentially as grid size is increased. Furthermore, ECOM is much faster than finite element based codes in terms of work per grid point, and requires less memory at equal grid size. Finally, ECOM can be easily parallelized in multi-core system since the solver on the unit disk uses separation of variables and solves an independent radial ODE for each angular grid point.

Because of the crowding effect inherent to the mapping of an elongated shape to a disk, ECOM usually requires a denser grid than FEM solvers to achieve the same accuracy. For tokamak geometries, this weakness is compensated by the speed of the solver: beyond a threshold grid size, ECOM computes the equilibrium quantities that play a key role in wave propagation, stability and transport calculations with more accuracy than CHEASE at equal run time. The threshold grid size depends on the plasma geometry and on the quantity that is calculated. For an ITER-like geometry and quantities that depend on first derivatives of the flux, such as the safety factor and the poloidal magnetic field energy, the threshold grid size is $n_{r}=n_{\vartheta}\simeq360$. For plasma shapes that are less elongated, the threshold grid size is smaller. It is also smaller if the quantity of interest depends on second derivatives of the flux, such as the local magnetic shear for instance, as a direct consequence of the exponential convergence of the second derivatives in ECOM. We also find the threshold grid to be smaller when the quantity of interest is a flux derivative of a flux function, such as the flux averaged magnetic shear for instance. This is because ECOM uses a Chebyshev grid for the flux contours and spectral differentiation for the evaluation of flux derivatives. 

ECOM has two important limitations. First, equilibria with $\kappa>3$ require too dense a grid for ECOM to reach a high level of accuracy in a reasonable amount of computing time. In these situations, particularly relevant to FRCs, ECOM is not an attractive option. Second, ECOM can only compute equilibria whose boundaries are smooth. It can therefore not be used for equilibria with a magnetic X-point. A promising idea to address these limitations is to develop a Grad-Shafranov solver based on an integral equation formulation that avoids conformal mapping to treat the geometrical aspects of the problem. Approaches relying on the Fast Multipole Method \cite{mckenney,ethridge} may represent an attractive option, that would lead to high order accuracy for the solution of the G-S equation as well as its derivatives. They are the subject of ongoing research. 

Note that there are additional options in ECOM that can be very desirable for certain applications, but are not discussed in this article. For instance, ECOM can handle equilibria that are specified in terms of the parallel current $J_\|(\psi)$ or the safety factor $q(\psi)$ instead of the poloidal current $F(\psi)$. Any one of the three profiles can be used in ECOM along with the specification of the pressure profile. When either the $J_\|$ profile or the $q$ profile constrains the equilibrium, ECOM needs to evaluate the flux functions $I_A$, $I_B$, and $I_C$ in Eq. (\ref{int123}) at each iteration. The fast and accurate numerical methods implemented in ECOM to calculate these functions then become a key strength of the solver, leading to fast convergence of the iterations and accurate equilibria. ECOM can also compute equilibria specified by an EFIT g-file \cite{laoEFIT} containing the pressure profile, the poloidal current profile, and the boundary shape of interest. And conversely, ECOM can print the results of any equilibrium calculation according to the format of an EFIT g-file, which includes the pressure, poloidal current and $q$ profiles, as well as the boundary and $\psi(R_E,Z_E)$ where $(R_E,Z_E)$ is a uniform grid in the range of $R_{min}<R_E<R_{max}$ and $Z_{min}<Z_E<Z_{max}$. For the sake of clarity and conciseness of the presentation, we did not describe these capabilities in the present article. They will be presented in detail in forthcoming articles, in which we explicitly use them to explore properties of tokamak equilibria and to couple ECOM with wave propagation and transport codes.  

\section*{Acknowledgments}

The authors would like to thank L. Greengard, E. Hameiri, M. O'Neil, and A. Pataki for helpful discussions, and M. O'Neil for the conformal mapping code. This research was supported in part by the U.S. Department of Energy, Office of Science, Fusion Energy Sciences under Award No. DE-FG02-86ER53223.
\appendix

\section{Miller parametrization of the flux contours}\label{sec:Miller}
Any flux contour of an up-down symmetric tokamak equilibrium can be approximated by a closed curve parametrized by the Miller parametrization given by Equations (\ref{millerb1})-(\ref{millerb2}) \cite{miller}, which we repeat below for convenience:
\begin{align}
R_M(t) &= R_{m0}+a \cos (t+\sin^{-1}\delta_m \sin t))\\
Z_M(t)&= a\kappa\sin (t)
\end{align}
ECOM has the option to compute the parameters $R_{m0}$, $a$, $\kappa$ and $\delta_m$ that provide a good approximation, in the least square sense, of a flux contour chosen by the user, and does this as follows. Given the numerical coordinates $(R_{i},Z_{i})_{i=1..n_{\theta E}}$ of the contour as a result of Sec. \ref{sec:post2}, ECOM calculates the vector $\mathbf{C}=[C_1, C_2, C_3]=[R_{m0}, a, \sin^{-1}\delta_m]$ that minimizes the sum $s$ of squared residuals
\begin{equation}
s=\frac{1}{2}\sum_{i=1}^{n_{\theta E}}(R_{M,i}-R_{i})^{2}
\end{equation}
where $R_{M,i}=R_M(t=t_i)$ and the parameter values $t_i$ are chosen so that $Z_m(t_{i})=Z_i$ for all $i$ in the integer interval $[1..n_{\theta E}]$. The product $a\kappa$ is held fixed during the minimization, and defines $\kappa$ once $C_2$ is calculated. The value $a\kappa$ is given by the condition $Z_m(\pi/2)=\mbox{max}_{i=1..n_{\theta E}}(Z_{i})$.

ECOM finds the vector $\mathbf{C}$ that minimizes $s$ by searching for the zero of $|\nabla s|$ with the Newton-Raphson method. Specifically, the sequence $\mathbf{C}^{(i)}$ of improved approximations of the minimizer $\mathbf{C}_{min}$ is given by
\begin{eqnarray}
\mathbf{C}^{(i+1)}=\mathbf{C}^{(i)}-\mathbf{\Omega}^{-1}\boldsymbol{\omega}
\label{lsbeta}
\end{eqnarray}
where $\mathbf{\Omega}^{-1}$ is the inverse of the Hessian matrix $\Omega$ defined by $(\Omega)_{ij}=\partial^{2}s/\partial C_{i}\partial C_{j}$, $\boldsymbol{\omega}=\partial s/\partial \mathbf{C}$ is the gradient vector, and the superscripts correspond to the iteration number. The iterative procedure stops when $|s^{(i+1)}-s^{(i)}|< \epsilon_{conv}$ is satisfied, for some prespecified $\epsilon$. For $\epsilon_{conv}=10^{-14}$, the convergence criterion is typically satisfied after $5-10$ steps. The components of the gradient vector are
\begin{equation}
\omega_{1}=\sum_{i}^{n_{\theta E}} (R_{M,i}-R_i),\;\;\omega_{2}=\sum_{i}^{n_{\theta E}} (R_{M,i}-R_i)\cos (\theta_{i}+C_3 \sin \theta_{i}),\;\;\omega_{3}=-C_2\sum_{i}^{n_{\theta E}} (R_{M,i}-R_i)\sin(\theta_{i}+C_3 \sin \theta_{i})\sin\theta_{i}\end{equation}

The entries of the Hessian matrix are
\begin{eqnarray}
\Omega_{11}&=&\sum_{i}^{n_{\theta E}} 1=n_{\theta E},\nonumber\\
\Omega_{12}&=& \Omega_{21}=\sum_{i}^{n_{\theta E}} \cos(t_{i}+C_{3}\sin t_{i})\nonumber\\
\Omega_{13}&=& \Omega_{31} =-C_{2}\sum_{i}^{n_{\theta E}}\sin t_{i}\sin(t_{i}+C_{3}\sin t_{i})\\ 
\Omega_{22}&=&\sum_{i}^{n_{\theta E}} \cos^{2}(t_{i}+C_{3}\sin t_{i}),\nonumber\\
\Omega_{23}&=&\Omega_{32}=-\sum_{i}^{n_{\theta E}} \left[C_{2}\cos(t_{i}+C_{3}\sin t_{i})+R_{M,i}-R_{i}\right]\sin t_{i}\sin(t_{i}+C_{3}\sin t_{i}),\nonumber\\
\Omega_{33}&=&C_{2}\sum_{1}^{n_{\theta E}}\sin^{2}t_{i}\left[C_{2}\sin^{2}(t_{i}+C_{3}\sin t_{i})-\cos(t_{i}+C_{3}\sin t_{i})(R_{M,i}-R_{i})\right]\nonumber
\end{eqnarray}
ECOM starts the iterative procedure with the following initial guesses
\begin{equation}
C_1=\frac{1}{n_{\theta E}}\sum_{i=1}^{n_{\theta E}}R_i,\qquad C_2=R_i|_{\mbox{max}(R_i)}-C_1,\qquad C_3=\frac{C_1-R_i|_{\mbox{max}(Z_i)}}{C_2}
\end{equation}
\pagebreak
\section{Summary of namelist variables in ECOM}\label{sec:namelist}

 \newcommand{\minrowheight}{\rule{0pt}{4ex}\relax}
\begin{longtable}{ ||  c  || c ||  c ||}
\caption {ECOM input variables in the Fortran namelist. Variables whose names start with 'I', 'N' or 'K' are in integer format, and those whose names start with 'file' are in string format. All other variables are in real format} \\
\hline
VARIABLE&\multicolumn{2}{c||}{DEFINITION}  \\
\hline
 &VALUE & DESCRIPTION \\
\hline
   IECOM& \multicolumn{2}{c||}{Profile specified along with the pressure profile in the G-S equation} \\ \cline{2-3}
            & 0 (default) & Poloidal current profile $d\bar{F}^2/d\psi$ \\ \cline{2-3}
           & 1& Parallel current profile $J_\|$  \\ \cline{2-3}
            &  2& Safety factor profile $q$  \\ \hline
   IPTYPE& \multicolumn{2}{c||}{ Specification of the pressure profile} \\ \cline{2-3}
            & 0 &  $\mu_0 d\bar{p}/d\psi=-C_s$ for the Solov'ev solution given by Eq. (\ref{solo2})\\ \cline{2-3}
           & 1 (default)&  $\mu_0d\bar{p}/d\psi=p_{0\psi}(1-(1-\psi)^{p_{in}})^{p_{out}}$ \\ \cline{2-3}    
            &  2& Discrete values of $d\bar{p}/d\psi$ or $p$ in terms of $\psi$ or $\rho$ are given by a table in `file\_prof'  \\ \cline{2-3}
           &3&  Discrete values of $d\bar{p}/d\psi$ in terms of $\psi$ is given by the EFIT output `file\_efit'  \\ \cline{2-3}
            & \multicolumn{2}{c||}{$C_s=F_B(\kappa+1/\kappa)/(R_0^3q_0^3)$ is determined by the namelist variables `F0',`q0',`rkappa', and `R0'}  \\ \cline{2-3}
           & \multicolumn{2}{c||}{The namelist variables for $p_{0\psi}$,${p_{in}}$ and ${p_{out}}$ are `p0psi', `pin' and ,`pout', respectively}  \\ \hline           
   IFTYPE& \multicolumn{2}{c||}{ Specification of the poloidal current profile (activated for IECOM=0) } \\ \cline{2-3}
            & 0 &   $d\bar{F}^2/d\psi=0$ for the Solov'ev solution given by Eq. (\ref{solo2})  \\ \cline{2-3}
           & 1(default)&         $(1/2)(d\bar{F}^2/d\psi)=F_{0\psi}(1-(1-\psi)^{f_{in}})^{f_{out}}$   \\ \cline{2-3}
           &  2& Discrete values of $(1/2)(d\bar{F}^2/d\psi)$ or $F$ in terms of $\psi$ or $\rho$ are given by a table in `file\_prof'  \\ \cline{2-3}
          &3&Discrete values of $(1/2)(d\bar{F}^2/d\psi)$ in terms of  $\psi$ given by the EFIT output `file\_efit' \\ \cline{2-3}      
           & \multicolumn{2}{c||}{The namelist variables for $F_{0\psi}$,${f_{in}}$ and ${f_{out}}$ are `ff0',`ffin' and ,`ffout', respectively}  \\ \hline       
   IJTYPE& \multicolumn{2}{c||}{ Specification of the parallel current profile (activated for IECOM=1) } \\ \cline{2-3}
           & 1(default)&         $J_\|=J_{\|0}(1-(1-\psi)^{j_{in}})^{j_{out}}$   \\ \cline{2-3}
           &  2& Discrete values of $J_\|$ in terms of $\psi$ are given by a table in `file\_jprof'  \\ \cline{2-3}
                 &3& $J_\|$ as evaluated from ohmic and bootstrap current models at each iteration\\ \cline{2-3}
           & \multicolumn{2}{c||}{the namelist variables for $J_{\|0}$,${j_{in}}$ and ${j_{out}}$ are `jpar0',`jpin' and ,`jpout', respectively}  \\ \hline    
  IQTYPE& \multicolumn{2}{c||}{Specification of the $q$ profile (activated for IECOM=2) } \\ \cline{2-3}
           & 1(default)&         $q=q_{0}(1+q_{fac}(1-\psi)^{q_{in}})^{q_{out}}$   \\ \cline{2-3}
           &  2& Discrete values of $q$ in terms of $\psi$ are given by a table in `file\_qprof' \\ \cline{2-3}
              & \multicolumn{2}{c||}{The namelist variables for $q_{0}$,${q_{fac}}$ $q_{in}$ and ${q_{out}}$ are `q0',`qfac',`qpin' and ,`qpout', respectively}  \\ \hline
 IBTYPE &  \multicolumn{2}{c||}{Specification of the plasma boundary $\partial\Omega$ } \\ \cline{2-3}
  & 0 & $\partial\Omega$ is given by Eqs. (\ref{solob1}-\ref{solob2}) \\ \cline{2-3}
    & 1 (default)& $\partial\Omega$ is given by Eqs. (\ref{millerb1}-\ref{millerb2}) \\ \cline{2-3}
 &2 &($R_n$,$Z_n$) are given by a table in 'file\_bc'  \\ \cline{2-3}
 &3 & ($R_n$,$Z_n$) are given by the EFIT output file `file\_efit' \\ \hline
 ITFTYPE &  \multicolumn{2}{c||}{ Specification of the toroidal flow pressure profile $\bar{p}_\Omega$} \\\cline{2-3}
             &0&No toroidal flow:  $\bar{p}_\Omega=0$\\\cline{2-3}
           &1&  $\bar{p}_\Omega=(M^2/2)\bar{p}_0$ where $M$ is the constant Mach number \\\cline{2-3}
            &2& $\bar{p}_\Omega=\bar{p}_{\Omega 0}(1-(1-\psi)^{p_{\Omega in}})^{p_{\Omega out}}$\\\cline{2-3}
            &3& Discrete values of $\bar{p}_\Omega$ in terms of $\psi$ are given by a table in `file\_tflow'  \\ \cline{2-3}
           & \multicolumn{2}{c||}{The namelist variables for $M$, $\bar{p}_{\Omega 0}$,$p_{\Omega in}$ and ${p _{\Omega out}}$ are `mach',`ptf0',`ptfin' and ,`ptfout', respectively}  \\ \hline    
   IPTABLE& \multicolumn{2}{c||}{ Type of pressure profile table in `file\_prof'  (activated for IPTYPE=2) } \\ \cline{2-3}
           & 0(default)&        1-D arrays of $\psi$ and $d\bar{p}/d\psi$ are given  \\ \cline{2-3}
           &  1&  1-D arrays of $\psi$ and ${p}$ are given  \\ \cline{2-3}
           &  2&  1-D arrays of $\rho$ and ${p}$ are given  \\ \hline
IRHO& \multicolumn{2}{c||}{ Definition for the normalized radius $\rho$  } \\ \cline{2-3}
           & 0(default)&       $\rho(\psi)=(R_o(\psi)-R_0)/(R_o(\psi=0)-R_0)$  \\ \cline{2-3}
           &  1&   $\rho(\psi)=({R_o(\psi)-R_i(\psi)})/({R_o(\psi=0)-R_i(\psi=0)})$  \\ \cline{2-3}
           &  2&  $\rho(\psi)=\sqrt{(\Psi-\Psi_0)/(\Psi_b-\Psi_0)}$.  \\ \hline
    nt1 &  \multicolumn{2}{c||}{ Number of grid points on the boundary used for the forward conformal mapping ($n_1$) } \\ \hline
    nt2 &  \multicolumn{2}{c||}{ Number of $\vartheta$ grid points in the unit disk ($n_\vartheta$) } \\ \hline
    nt3 &  \multicolumn{2}{c||}{ Number of $\theta$ grid points for contour integrals ($n_{\theta D}=n_{\theta E}$)} \\ \hline  
    nsub &  \multicolumn{2}{c||}{ Number of radial piecewise Chebyshev intervals in the unit disk ($n_L$)} \\ \hline
    kcheb &  \multicolumn{2}{c||}{ Number of Chebyshev points in a radial interval ($n_{ch}$)} \\ \hline
    nchq &  \multicolumn{2}{c||}{ Number of flux surfaces in a Chebyshev grid of $\psi$ for postprocessing ($n_f$)} \\ \hline
    nflx &  \multicolumn{2}{c||}{ Number of flux surfaces in uniform grid of $\rho$ for postprocessing} \\ \hline
    kLag &  \multicolumn{2}{c||}{ Order of Lagrange interpolation ($k_{Lag}$)} \\ \hline
    ksamp &  \multicolumn{2}{c||}{ Oversampling factor for FFT padding used for contour integrals ($k_{samp}$)} \\ \hline
    R0  &  \multicolumn{2}{c||}{ R coordinate of the point that is mapped to the center of $D_1$ by the initial conformal mapping} \\ \cline{2-3}
& \multicolumn{2}{c||}{ $R_0$ for IBTYPE=0 and $R_{m0}$ for IBTYPE=1} \\ \hline
    Z0    &  \multicolumn{2}{c||}{ Z coordinate of the point that is mapped to the center of $D_1$ by the initial conformal mapping} \\ \cline{2-3}
& \multicolumn{2}{c||}{ $Z_0$ for IBTYPE=0 and $Z_{m0}$ for IBTYPE=1} \\ \hline
    q0  &  \multicolumn{2}{c||}{ Value of the safety factor at the magnetic axis $q(\Psi=\Psi_0)$} \\ \hline
    F0  &  \multicolumn{2}{c||}{ $F(\Psi=\Psi_0)$ for IFPOL=0 or $F(\Psi=\Psi_B)$ for IFPOL=1 } \\ \hline
    ISCALE &  \multicolumn{2}{c||}{ Control parameter to scale the solutions. See Table \ref{tab:iscale} for further details} \\ \hline 
    torcur &  \multicolumn{2}{c||}{ Total toroidal current in [MA] for ISCALE=2 or ISCALE=3} \\ \hline
    reps &  \multicolumn{2}{c||}{ Ratio of minor radius to major radius. $a/R_{0}$ for IBTYPE=0 and $a/R_{m0}$ for IBTYPE=1} \\ \hline
    rkappa &  \multicolumn{2}{c||}{ Elongation of the boundary. $\kappa$ for IBTYPE=0 and IBTYPE=1} \\ \hline
    delta  &  \multicolumn{2}{c||}{ Triangularity of the boundary. $\delta_m$ for IBTYPE=1} \\ \hline
    epsiter &  \multicolumn{2}{c||}{ Small constant to determine the convergence of iteration ($\delta$)} \\ \hline
    nmaxiter &  \multicolumn{2}{c||}{ Maximum number of iterations} \\ \hline
    epsmaxdist &  \multicolumn{2}{c||}{ Maximum distance between mapping center and the magnetic axis} \\ \hline
    ISYMUD &  \multicolumn{2}{c||}{ Index for up-down symmetry of $\Psi$ and boundary about Z=0 axis  (0: asymmetric, 1: symmetric)} \\ \hline
     IPRINTSOL     & \multicolumn{2}{c||}{  Printing an Ascii file of $\Psi$ and its derivatives on the grid of $\Omega$ (0: off, 1: on)} \\ \hline
    IPRINTMAP     & \multicolumn{2}{c||}{  Printing Ascii files for conformal mapping results (0: off, 1: on)} \\ \hline
    IPRINTSOLDISK &  \multicolumn{2}{c||}{  Printing an Ascii file of $\Psi$ and its derivatives on the grid of the unit disk $D_1$ (0: off, 1: on)} \\ \hline
    IPRINTCON     & \multicolumn{2}{c||}{ Printing Ascii files of contours of nchq $\psi$ values and nflx $\rho$ values (0: off, 1: on)} \\ \hline
    IPRINTQS      &  \multicolumn{2}{c||}{ Printing an Ascii file of the safety factor and the magnetic shear in terms of nchq $\psi$ and nflx $\rho$  (0: off, 1: on)} \\ \hline
     IPRINTEFIT   &  \multicolumn{2}{c||}{ Printing an Ascii file in EFIT g file format (0: off, 1: on)} \\ \hline
      npsi    &  \multicolumn{2}{c||}{ Number of flux surfaces for EFIT g file format for iprintefit=1} \\ \hline 
       ISTABILITY      &  \multicolumn{2}{c||}{ Evaluation of the Mercier criterion and Troyon limit (0: off, 1: on)} \\ \hline
       IBSCUR     &  \multicolumn{2}{c||}{ Evaluation of the bootstrap and ohmic currents (0: off, 1: on)} \\ \hline    
       IJBSMODEL    &  \multicolumn{2}{c||}{ Bootstrap and ohmic current model (1: Hirshman model, 2: Sauter model)} \\ \hline 
       nchy    &  \multicolumn{2}{c||}{ Number of pitch angle grid points to evaluate the bootstrap and ohmic currents } \\ \hline 
       VLOOP0  &  \multicolumn{2}{c||}{ Loop voltage giving the ohmic current in unit of volt}\\ \hline 
      IFITMIL   &  \multicolumn{2}{c||}{ Fitting the flux surfaces using the Miller parametrization (0: off, 1: on)} \\ \hline    
\hline
\label{tab:pinch_diff}
\end{longtable}

\section*{References}


\end{document}